\documentclass[conference]{IEEEtran}
\IEEEoverridecommandlockouts

\usepackage[labelformat=simple]{subcaption}

\usepackage[linesnumbered,ruled,vlined]{algorithm2e}
\usepackage{bm}
\usepackage{bbm}
\usepackage{glossaries}
\usepackage{mathtools}
\usepackage{cite}
\usepackage{amsmath,amssymb,amsfonts}
\usepackage{algorithmic}
\usepackage{graphicx}
\usepackage{textcomp}
\usepackage{comment}
\usepackage{multirow}

\def\BibTeX{{\rm B\kern-.05em{\sc i\kern-.025em b}\kern-.08em
    T\kern-.1667em\lower.7ex\hbox{E}\kern-.125emX}}

\newtheorem{definition}{Definition}
\newtheorem{example}{Example}
    
\makeglossaries

\newglossaryentry{ppp}
{
    name=\textit{Pivot Permutation Prefix},
    description={}
}

\newglossaryentry{pp}
{
    name=\textit{Pivot Permutation},
    description={}
}

\newglossaryentry{infra}
{
    name=CLIMBER,
    description={}
}

\newglossaryentry{index}
{
    name=CLIMBER-INX,
    description={}
}

\newglossaryentry{p4}
{
    name=$P^4$ signature,
    description={}
}

\newglossaryentry{p4s}
{
    name=$P^{4\rightarrow}$,
    description={}
}

\newglossaryentry{p4i}
{
    name=$P^{4\nrightarrow}$,
    description={}
}

\newglossaryentry{is}
{
    name={\em INX-IS},
    description={}
}

\newglossaryentry{trie}
{
    name={\em INX-Trie},
    description={}
}

\begin{document}
\title{CLIMBER++: Pivot-Based Approximate Similarity Search over Big Data Series}

\author{
{\fontsize{11}{11}\selectfont
Liang Zhang$^+$, ~~~Mohamed Y. Eltabakh$^{\star}$,~~~Elke A. Rundensteiner$^{\dagger}$,
Khalid Alnuaim$^{\dagger}$}
 \\
\fontsize{10}{10}\selectfont\rmfamily\itshape
$^+$ Oracle Inc., $^{\star}$ Qatar Computing Research Institute (QCRI), 
$^{\dagger}$ Worcester Polytechnic Institute (WPI)
\\
\fontsize{9}{9}\selectfont\ttfamily\upshape
\{lzhang@Oracle.com, meltabakh@hbku.edu.qa,  rundenst@wpi.edu, kalnuaim@wpi.edu\}
}

\maketitle

\begin{abstract}
The generation and collection of big data series are becoming an integral part of many emerging applications in sciences, IoT, finance, and web applications among several others.
The terabyte-scale of data series has motivated recent efforts to design fully distributed techniques for supporting operations such as {\em approximate kNN similarity search}, which is a building block operation in most analytics services on data series. Unfortunately, these techniques are heavily geared towards 
achieving scalability at the  cost of
sacrificing the results' accuracy.
State-of-the-art systems DPiSAX and TARDIS report accuracy below 10\% and 40\%, respectively, which is not practical for many real-world applications. 
In this paper, we investigate the root problems in these existing techniques
that limit their ability to achieve better a trade-off between scalability and accuracy. Then, we propose a framework, called CLIMBER, 
that encompasses a novel feature extraction mechanism, indexing scheme, and query processing algorithms for supporting approximate similarity search in big data series. 
For  CLIMBER, we propose a new loss-resistant dual 
representation composed of 
 {\em rank-sensitive} and {\em ranking-insensitive} signatures capturing data series objects. Based on this representation, we devise a distributed two-level index structure supported by an efficient data partitioning scheme. Our similarity metrics tailored for this dual representation 
enables meaningful comparison and distance evaluation between the  {\em rank-sensitive} and {\em ranking-insensitive} signatures. 
Finally, we propose two efficient query processing algorithms,   
    {\em CLIMBER-kNN} and  {\em CLIMBER-kNN-Adaptive},  
    for answering approximate kNN similarity queries.  
Our experimental study
on  real-world and benchmark datasets 
demonstrates that CLIMBER,
unlike existing techniques, features results' accuracy above 80\% while 
retaining the desired scalability to terabytes of data. \footnote{This paper is an extended version of the ICDE 2024 manuscript.}
\end{abstract}

\section{Introduction}
\label{sec:Intro}
Many emerging applications in science, engineering, IoT, and social 
generate data series at an explosive speed \cite{chan2004time,palpanas2016big, cook2019anomaly, aljawarneh2016similarity, carrington2005models, wei2006time}. 
For example, an ECG (electrocardiogram) device generates data series of approximately $1$ gigabyte per hour, 
a typical weblog tracing generates around $5$ gigabytes per week, 
and a space shuttle generates data series of roughly $2$ gigabytes per day~\cite{aghabozorgi2015time}.
As a result, numerous mining algorithms, e.g., clustering~\cite{aghabozorgi2015time}, classification~\cite{abanda2019review}, motif discovery~\cite{torkamani2017survey}, outlier patterns~\cite{abraham1989outlier}, and segmentation~\cite{sclove1983time},
on such rich data series repositories have been developed--with 
similarity search being a critical building block operation in all of these algorithms~\cite{esling2012time,jensen2017time,palpanas2017parallel}.

Given that a data series object is typically high dimensional, e.g., 
a single data series may contain 100s of readings, 
similarity search algorithms have  relied on 
indexing techniques to avoid prohibitively expensive 
full scans, and thus achieve better efficiency and 
scalability~\cite{palpanas2017parallel,echihabi2019return, shieh2008sax, camerra2010isax, zhang2019tardis, Odyssey147779087, 10.1Messi00677-2}.
Intrinsic to these indexing techniques are the 
{\em feature extraction} and {\em dimensionality reduction} phases 
that first create a feature representation for the data in a lower dimensional space, and then construct an index
on this lower dimensional representation, e.g., 
~\cite{faloutsos1994fast, chan1999efficient, keogh2001dimensionality, lin2007experiencing, shieh2008sax}.
Undoubtedly, both the feature extraction technique
and the index structure 
play a critical role 
in deciding the results' accuracy 
of the similarity search operator.

Focusing on approximate similarity search on big data series,   the scope of our work, few 
of these aforementioned systems offer distributed indexing solutions 
capable of scaling to terabytes of data, e.g.,~\cite{zhang2019tardis,yagoubi2017dpisax,zhang2020big, palpanas2017parallel, mottin2019exploring}. 
However, in these techniques, scalability is achieved at the cost
of sacrificing the results' accuracy.  
In many cases, the accuracy reported about the state-of-the-art 
techniques~\cite{zhang2019tardis,yagoubi2017dpisax,zhang2020big} 
is too low to be reliable for most applications--with a maximum achieved accuracy around 40\% as reported in ~\cite{zhang2019tardis, echihabi2019return}. 
This is because existing techniques are heavily geared towards 
improving the query's speed and scalability~\cite{palpanas2017parallel, mottin2019exploring,echihabi2019return} over improving the result's accuracy.

\vspace{1mm}
Even more critically, we observe that there are 
fundamental limitations in the existing techniques that prevent 
achieving notable higher accuracy while retaining the desired scalability. 
For example, the performance and accuracy of R-tree indexes, 
which are built on top of representations such as DFT, DWT, PAA and SAX,
are known to degrade quickly under high-dimensional data 
due to the curse of dimensionality~\cite{bellman1966dynamic}. 
The more recent iSAX representation 
relies on dividing a data series into equal-size segments and calculating 
the mean value of each segment as its representation~\cite{shieh2008sax, zhang2019tardis} 
(see Figure~\ref{fig:sax-isax-rep}). 
But under higher dimensionality, either the number of segments 
is  kept small but then the segment's length increases, or vice versa. The former choice results in a too lossy representation 
while the latter choice results in slower index construction and query processing.   
Existing indexes on iSAX, e.g., 
iSAX-Binary~\cite{camerra2010isax} and 
{\em sigTree}~\cite{zhang2019tardis}, opt for the 
former  because otherwise both the query speed and indexing scalability 
degrade  quickly. 

Based on the aforementioned observations, 
we put further  that 
the sweet spot 
for the trade-off between speed and accuracy over large-scale data series remains elusive. 
Therefore, there is a critical need for a novel  framework that succeeds in solving two main challenges concurrently: 
(1) a feature extraction approach that is loss-resistant 
under high dimensionality, and 
(2) an indexing mechanism that preserves its scalability and search performance 
under longer feature representations.
In this paper, we propose such a framework, called ``\gls{infra}'', effectively supporting approximate similarity queries  
over high dimensional data series.
\gls{infra} boosts the results' accuracy to an unprecedented
level close to 80\% while still scaling well to 100s of dimensions and terabyte-scale of data. \footnote{CLIMBER source code: https://github.com/lzhang6/climber.git}

The feature extraction mechanism in \gls{infra}, referred to as \gls{infra}-FX,
utilizes the {\em pivot-permutation-prefix} technique in~\cite{novak2016ppp,esuli2012use}
for creating a novel dual representation for data series objects. 
More specifically, \gls{infra}-FX divides the data space into  Voronoi partitions, each having its own pivot. 
Then, each data series, depending on its location within 
the space, is represented by 
dual pivot vectors, namely  {\em rank-sensitive} and {\em rank-insensitive} representations. 
The former is an ordered list of pivots based on the proximity to the data series object, 
while the latter is an ordered list of pivots based on a global ordering scheme 
(e.g., lexicographical order) independent of 
the proximity to the data series object. 
Leveraging this dual  representation scheme, we design
 a hybrid index structure, 
referred to as \gls{infra}-INX), that first clusters the 
data series into groups based on their similarity using the {\em rank-insensitive} representation,
and then further refines the groups into partitions using the 
{\em rank-sensitive} representation. 

For efficient index construction, we propose a data-driven sampling method
for discovering the centroids of the data groups. 
We show that the proposed dual pivot representation 
poses challenges to existing similarity measures as they are not designed  
for handling two different representations for each object in the same system. 
Therefore, we propose new similarity metrics customized to
this   
dual representation.
Lastly, we devise two  algorithms, 
{\em CLIMBER-kNN} and {\em CLIMBER-kNN-Adaptive}, 
for the efficient distributed evaluation of kNN queries.

In summary, the contributions of this paper are as follows:

\begin{itemize}

\item Presenting CLIMBER, a holistic system  that encompasses 
a novel feature extraction mechanism, indexing scheme, and query processing algorithms supporting approximate similarity search in big data series. Unlike existing techniques that primarily
focus on scalability, CLIMBER treats both scalability and accuracy as first-class citizens %and enables
boosting the 
accuracy of query results to 80\%.

\item Proposing a novel dual pivot-prefix representation composed of  
  {\em rank-sensitive} and {\em ranking-insensitive} 
  signatures.  Based on this,  we devise a two-level
  index structure that organizes the data into groups ($1^{st}$ level) 
  and then into partitions ($2^{nd}$ level). 
  Both levels consistently follow the pivot-based Voronoi partitioning scheme for seamless 
  index traversal.
  
  \item Introducing new similarity metrics tailored for the dual pivot representations. 
    They enable meaningful comparison and distance evaluation between the  {\em rank-sensitive} and {\em ranking-insensitive} signatures across the index layers.

  \item Proposing two query processing 
  algorithms,   
    {\em CLIMBER-kNN} and  {\em CLIMBER-kNN-Adaptive},  
    for efficient answering of approximate kNN similarity queries.  
    The latter adaptive solution identifies when the 
    best partition identified by {\em CLIMBER-kNN} may contain less than $k$ 
    high-quality answers and automatically adapts by expanding the search space.

  \item Developing a fully distributed prototype of CLIMBER that scales to terabyte data series repositories with 100s of dimensions. We conducted extensive experiments on benchmark and real-world datasets.
  The results are promising and show that CLIMBER achieves 
  the two desired yet competing properties of scalability and results' accuracy within the same system.
  
\end{itemize}

The rest of this paper is organized as follows. 
We present related work in Section~\ref{sec:Relwork}, and 
then review the background in Section~\ref{sec:preliminary}. 
The main  components of CLIMBER  of 
feature extraction (\gls{infra}-FX) and indexing (\gls{infra}-INX)  
are presented in Sections~\ref{sec:FX} and~\ref{sec:idx}, respectively. 
The kNN approximate query processing algorithms
(\gls{infra}-kNN, \gls{infra}-kNN-Adaptive) are described in Section~\ref{sec:query}. 
The experimental evaluation is given in Section~\ref{sec:exper}. 
Finally, we conclude in Section~\ref{sec:conclusion}.

%================================

\section{Related Work}
\label{sec:Relwork}

Data series similarity search has been extensively studied in literature from various aspects. 
We categorize the work related to our system as follows:

{\em Approximate vs. Exact Similarity Search:}
Exact similarity search over big data series repositories is very expensive and highly resource demanding. 
That is why approximate similarity search, e.g.,~\cite{zhang2019tardis, echihabi2019return, yagoubi2017dpisax, nourachainlink, wu2019kv, zhang2020big, isaxfamily, arora2018hd},  tends to be more practical   
for many applications, e.g., analytical, visualization, and exploration tasks. 
Recent systems such as Messi~\cite{10.1Messi00677-2} and its multi-node distributed version Odyssey~\cite{Odyssey147779087} have made it feasible to efficiently perform exact similarity search over large-scale datasets. 
Nevertheless, these systems are main-memory processing engines  
tailored to streaming and batch-query applications. 
Hence, they require the entire dataset and the constructed indexes to fit 
in main memory in order to deliver competitive performance.  
As a result, given the same compute resources, distributed disk-based systems 
such as CLIMBER can scale 
to much bigger datasets compared to the memory-based systems.

{\em Feature Extraction \& Indexing Mechanims:}
Feature extraction and indexing are intrinsic mechanisms to data series similarity search. Numerous feature extraction techniques have been proposed including 
DFT~\cite{faloutsos1994fast}, 
DWT~\cite{chan1999efficient}, 
wavelets~\cite{chan1999efficient},
PAA~\cite{keogh2001dimensionality}, 
SAX~\cite{lin2007experiencing}, 
and 
iSAX~\cite{shieh2008sax}. 
Associated with each representation, 
various index structures have been designed 
ranging from traditional Spatial Access Methods (SAMs) such as the R-tree~\cite{guttman1984r} or its variants~\cite{arge2008priority,beckmann1990r,kamel1993hilbert} to 
custom-designed indexes such as 
iSAX Binary Tree~\cite{camerra2010isax}, 
{\em TARDIS}~\cite{zhang2019tardis},  ChainLink~\cite{nourachainlink}, and Messi~\cite{10.1Messi00677-2, 9101877Messi}.

All recent systems, e.g.,~\cite{keogh2001dimensionality, lin2007experiencing, zhang2019tardis, shieh2008sax, nourachainlink, 10.1Messi00677-2, 9101877Messi, Odyssey147779087, zhang2020big, yagoubi2017dpisax}, rely on the data series-specific representations such as PAA, SAX, and iSAX as opposed to the traditional representations of DFT, DWT, and wavelets. This is not only because the former methods show superior performance with 
respect to index construction time and query accuracy, but also they are more flexible (e.g., they allow for queries shorter than the length on which the index is built, which is not possible in DFT and wavelet methods)~\cite{keogh2001dimensionality, 105555645803669368, lin2007experiencing}. 
Despite that, and as highlighted in Section~\ref{sec:Intro}, 
these systems fail to achieve the desired trade-off between scalability and accuracy.
The proposed CLIMBER system is a one-step forward towards bridging this gap and achieving a boost in the accuracy without sacrificing scalability.

{\em Hashing- and Graph-based Nearest Neighbor Search:}
Approximate nearest neighbor search (ANNS) is a core operation in various domains. 
With the rise of vector databases~\cite{Chroma333, PostgreSQL}, embedding-based similarities~\cite{DBLP:journals/corr/JohnsonDJ17}, and document-based cleansing for Large Language Models (LLMs)~\cite{RedPajama, Dolam23}, 
several kNN query algorithms for high-dimensional data have gained 
recent attention such as Locality Sensitive Hashing (LSH)~\cite{gh87287, 10.147777374} and Proximity Graph (PG)~\cite{10.110ww89473, 10.555555520, 1014562753538475}. 

In our previous work~\cite{nourachainlink}, we investigated leveraging LSH in the context of terabyte-scale data series repositories and presented ChainLink as the state-of-the-art technique. However, while ChainLink improves the query response time by 10x fold over the vanilla LSH algorithms~\cite{DBLP:conf/nips/LuoS16}, 
ChainLink shares the same limitation of the aforementioned techniques which is the low results' accuracy, i.e., recall is around 30\%.
This is primarily because LSH techniques rely on the potential of 
syntactic similarity among similar objects. 
Therefore, in the data series context with numeric values, lossy sketching techniques need to be first applied on the data before performing the hashing.

Graph-based methods, on the other hand, have recently shown very high accuracy (reaching  90\% and higher) 
in answering approximate kNN queries over high-dimensional vector databases. 
Examples of these techniques 
include DiskANN~\cite{10.555555520}, HNSW~\cite{10.110ww89473}, and  PyNNDescent~\cite{PyNNDescent}. Despite 
such remarkable accuracy, most of these techniques suffer from 
very heavy computational costs to construct the similarity graphs. 
For example, DiskANN reported a construction time of 7 days over a billion-scale dataset~\cite{10.555555520}. 
Moreover, graph-based techniques are extremely challenging to distribute due to their shared main-memory intensive computations and the locking mechanisms used in several of these techniques~\cite{10.110ww89473, PyNNDescent}. 
The recent work in ParlayANN~\cite{1014562753538475} 
introduced a multi-core shared-memory framework for parallelizing several graph algorithms in a single-node setting. 
During the construction phase, the entire graph and vector data need to fit 
in this shared memory. 
Even with such parallelization, ParlayANN-HNSW took 16 Hours 
to construct the graph  over a one billion vector.
To the best of our knowledge, there is no highly distributed 
multi-node graph-based method for kNN queries.

In summary, within the landscape of the 
state-of-the-art techniques presented above, CLIMBER lands itself in a spot of high demand by modern applications and yet lacking proper coverage. More specifically, scaling to big data that potentially far exceed the available main memory while achieving high accuracy and few-seconds query response time.

%=================================
\section{Preliminary}
\label{sec:preliminary} 

\subsection{Basic Terminology} 
\begin{definition} 
  \textbf{[Data Series]}
  \textit{
    A data series $X=[ x_{1}, x_{2},$ $\cdots, x_{n}]$ $x_{i} \in R$ is an ordered sequence of real-valued variables, with $|X| = n$ denoting the length of a data series.
 }
  \label{def:data-series}
\end{definition}
A data series of length $n$ is an object in  an 
$n$-dimensional space, where the $i^{th}$ reading is the value of dimension $i$.

\vspace{1mm}
\begin{definition} 
  \textbf{[Data Series Dataset]}
  \textit{
    A data series dataset $DB= \lbrace X_{1},X_{2},\cdots,X_{d} \rbrace$ is a collection of $d$ data series, each of the same length $n$.}    
\end{definition}

A data series distance is a function that measures the (dis)similarity of two data series. The Euclidean distance is the most widely used and one of the most effective for large series collections~\cite{ding2008querying}.

\vspace{1mm}
\begin{definition} 
 \textbf{[Euclidean Distance]} 
 \textit{
  Given two data series $X=[ x_{1}, x_{2}, \cdots, x_{n}]$ 
  and $Y=[ y_{1}, y_{2}, \cdots, y_{n}]$, their Euclidean distance is:}
 \begin{equation} 
  {ED(X,Y) = \sqrt{\sum\limits_{i=1}^n (x_{i} - y_{i})^2} } 
 \end{equation}\label{def:ed}
\end{definition}

\vspace{1mm}
\begin{definition}
  \textbf{[kNN Approximate Query]} 
  \textit{
    Given a data series dataset $DB$, a query data series $X_q$, and an integer value $k$, the query finds set $S_{approx}$ containing $k$ data objects that are the closest or potentially very close to $X_q$. Assuming hypothetically that the ground truth exact answer set $S_{exact}$ is available, the error in $S_{approx}$ is computed as:
    \begin{equation}
    {recall = \frac{|S_{approx}\cap S_{exact}|}{|S_{exact}|}}
    \end{equation}
}
  \label{def:knnapprox}
  \vspace{-1mm}
\end{definition}

{\em Recall} is popular standard measure used to represent the approximation accuracy, i.e., how far the reported answer set $S_{approx}$ is from the exact answer set $S_{exact}$~\cite{park2015neighbor, nourachainlink, zhang2019tardis}.

\subsection{State-of-the-Art Representation and its Limitations}
\label{sec:art-limitations}
\begin{figure}[t]
  \centering
  \begin{subfigure}[t]{0.49\columnwidth}
      \centering
      \includegraphics[width=1.0\textwidth]{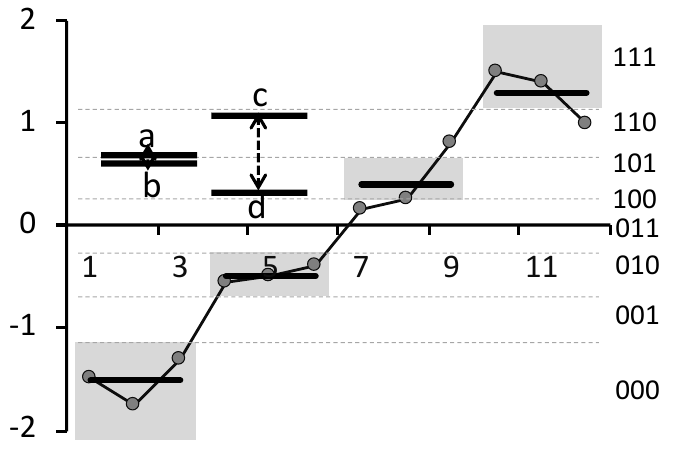}
      \caption{SAX = [$000$, $010$, $101$,$111$] }
      \label{fig:sax-rep}
  \end{subfigure}%
  ~
  \begin{subfigure}[t]{0.49\columnwidth}
    \centering
    \includegraphics[width=1.0\textwidth]{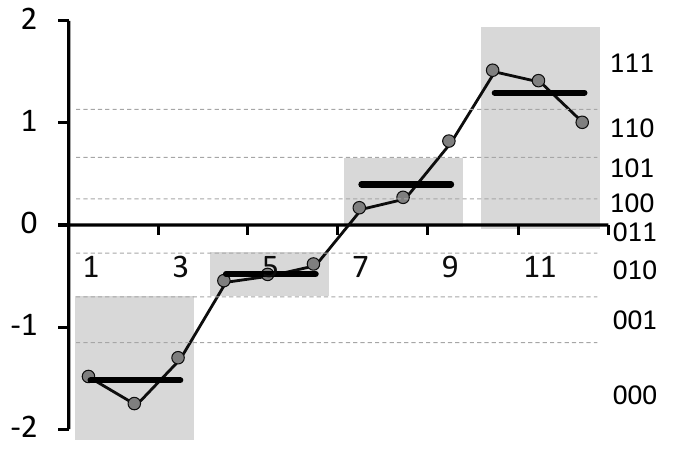}
    \caption{iSAX = [$00_2$, $010_3$, $10_2$, $1_1$]}
    \label{fig:isax-rep}
  \end{subfigure}%

  \caption{Examples of SAX and iSAX Repr. ($w=4, c=8$).}
  \label{fig:sax-isax-rep}
  \vspace{-4mm}
\end{figure}

The state-of-the-art big data series indexes are based on SAX and iSAX representations~\cite{yagoubi2017dpisax,zhang2019tardis}. 
The intuition of the SAX representation (Figure ~\ref{fig:sax-isax-rep}) 
is to divide a  data series object over the x-axis (i.e., the {\em time/order} dimension) 
into equal-length segments, 
where the total number of segments is referred to as the {\em ``word length" (w)}. 
Similarly, we divide the y-axis (i.e., the {\em value} dimension) into horizontal stripes, where
the number of these stripes (called {\em ``cardinality" ($c$)}) 
is typically a power of two. 
For example, the data series in Figure ~\ref{fig:sax-isax-rep}(a) is divided into four 
segments ($w=4$) over the x-axis, each consisting of three values,  
and into eight stripes ($c=8$) over the y-axis, each 
encoded with a unique 3-bit binary representation. 

The authors in~\cite{lin2007experiencing} proposed an algorithm to determine  the boundaries of each stripe.
For example, in Figure~\ref{fig:sax-rep}, stripes ``111'' and ``011'' have 
the boundaries of [1.15, $\infty$], and [-0.31, 0], respectively.  
A data series is then transformed into its corresponding SAX representation by 
calculating the mean value of each segment (illustrated by a 
horizontal line in Figure~\ref{fig:sax-rep}) and then depending on the 
stripe containing the mean value, the segment is  assigned its corresponding binary label. 
In the SAX representation, all segments use the same cardinality, e.g.,
the three-bit representation as depicted in Figure~\ref{fig:sax-rep}. 
Whereas, in the iSAX representation, the cardinality for different segments can  differ
and only prefixes are maintained as  depicted in Figure ~\ref{fig:sax-isax-rep}(b). 

Obviously, SAX-based representation involves two levels of lossy transformations. 
The first converts segments into their mean values, and the second 
assigns the mean values to stripe labels. 
For example, referring to Figure~\ref{fig:sax-rep}, 
segments $a$ and $c$ belong to the same stripe, i.e., $110$, 
and $b$ and $d$ also belong to the same stripe, i.e., $101$.
Hence, the distance between $a$ and $b$ is considered the same as 
that between $c$ and $d$ -- even though  they may
differ significantly in practice.
Such transformation loss may result in a weak similarity preservation during 
the similarity search across the data set. 

To index the iSAX representation,  tree-based index structures have been proposed 
including iBT~\cite{shieh2008sax, yagoubi2017dpisax} and sigTree~\cite{zhang2019tardis} indexes.
Both indexes are designed for distributed storage and parallel processing on infrastructures, e.g., Apache Hadoop and Spark.
 The two index structures have fundamental differences with respect to their structural design, e.g., the fanout size, the balancing properties, 
and the trade-offs between the tree width and depth. 
Both index structures tend to prefer a small word length $w$ (the number of segments)  to construct a compact tree with short traversal depths. 
A consequence of such small $w$ is that both index structures inherit low similarity preservation due to  
the lossy iSAX representations.  
In short, these systems~\cite{yagoubi2017dpisax,zhang2019tardis} 
sacrifice accuracy  to achieve scalability -- 
with the highest reported accuracy being 40\% in~\cite{zhang2019tardis}).

%=================================================================
%===============================================================

\section{CLIMBER Feature Extraction}
\label{sec:FX}

In this section, we present the 
feature extraction technique in \gls{infra}. The technique builds 
on two core concepts, namely, {\em pivot permutation}~\cite{amato2008approximate,gonzalez2008effective,chavez2005proximity} and 
{\em pivot permutation prefix}~\cite{novak2016ppp,esuli2012use}. 
In Section~\ref{sec:ppp-overview}, we explain these concepts along with their limitations when directly applied to the data series domain, 
while in Section~\ref{sec:sign}. we introduce our  \gls{infra}-FX solution.

%===========================
\subsection{Overview on Pivot Permutation Techniques}
\label{sec:ppp-overview}
\begin{figure}[t]
  \centering
  \includegraphics[width=0.99\columnwidth]{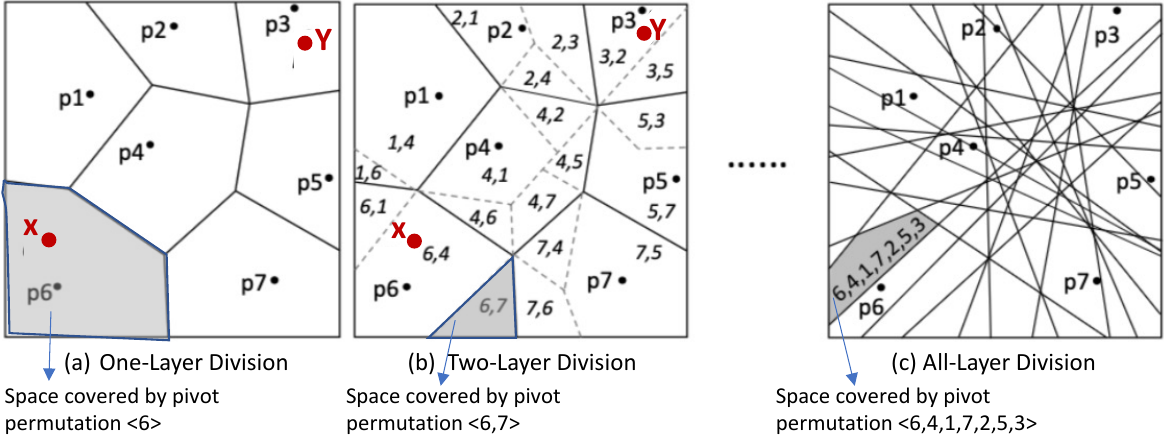}
  \caption{Recursive Voronoi partitioning. Fragments are labeled by the corresponding pivot or the sequences of pivots.}
  \label{fig:voronoi}
  \vspace{-4mm}
\end{figure}

The Pivot Permutation-based algorithm~\cite{amato2008approximate,gonzalez2008effective,chavez2005proximity} 
is an approximate retrieval method in the metric space  
where the distance function satisfies the metric postulates 
of {\em non-negativity, identity, and triangle inequality}~\cite{samet2006foundations}. 
A group of randomly selected {\em pivots} are used as reference objects to divide the space into Voronoi partitions as illustrated in Figure~\ref{fig:voronoi}.
Then, the pivot permutations correspond to a recursive Voronoi-like partitioning of the metric space~\cite{skala2009counting}.
For example, in Figure~\ref{fig:voronoi}(a),  seven pivots ($p_1$ to $p_7$) are used 
to divide the given Euclidean space into Voronoi cells. 
Then, the dotted lines in Figure~\ref{fig:voronoi}(b) 
further partition these Voronoi cells into two-layer partitions.
For example, the partition marked with {\em ``6-7''}  
indicates that any point in this partition has Pivots $p_6$ followed by $p_7$ as 
the closest pivots. 
Figure~\ref{fig:voronoi}(a) is the simplest one-layer 
Voronoi partitioning of the space 
while Figure~\ref{fig:voronoi}(c) is the full seven-way recursive Voronoi partitioning.

Given a data object $X$, the distances between $X$ and all pivots are computed, and then the pivots are sorted in an ascending order to create $X$'s {\em pivot permutation}, which is basically $X$'s representation in the feature space.  
For example, data point $X$ has the associated pivot permutation of 
$<p_6$,$p_4$,$p_1$,$p_7$,$p_2$,$p_5$, $p_3>$, and hence 
its corresponding representation is  ``$<6,4,1,7,2,5,3>$''. 
Similarly, data point $Y$  has the representation of 
``$<3,2,5,4,1,7,6>$''.
For a given similarity query $X_q$, the search algorithm 
starts by converting $X_q$ to its pivot permutation representation, 
and then searches for data objects whose representation 
is similar to that of $X_q$. 
The assumption is that if two objects are very close to one another, 
then their pivot permutations will be very similar in the feature space. 

Given that the number of pivots can be large, the vector representation for each data point can be long. Thus it consumes considerable space as well as expensive computations during comparisons. 
To overcome these limitations, the  \gls{ppp} 
technique  has been proposed~\cite{novak2016ppp,esuli2012use}.

\vspace{2mm}
\begin{definition}\label{def:ppp}
  \textbf{[Pivot Permutation Prefix]}
  \textit{ 
    Given a set of $r$ pivots $\{p_1,p_2,$ $p_3,\cdots,p_r\} \subseteq \mathcal{D}$, a metric distance \textbf{md}, an object \textbf{o} $ \in \mathcal{D}$ 
    and a prefix length 
    $m \leqslant r$, the Pivot Permutation Prefix of \textbf{o} denoted as 
    {\textbf{PPP(o)}} = $<p^o_{\Pi_1}, p^o_{\Pi_2}, \cdots, p^o_{\Pi_m}>$
    is an ordered vector of the $m$ nearest pivots to \textbf{o}, where  
    $p^o_{\pi_i}$ indicates the pivot at position $i$ and }
    {\small
      \begin{displaymath}
        md(p^o_{\pi_i},o) <  md(p^o_{\pi_{i+1}},o), ~\forall i: 1 \leqslant i \leqslant m 
      \end{displaymath}
    }
\end{definition}

As an example, Figure~\ref{fig:voronoi}(b) illustrates a recursive Voronoi partitioning of $depth=2$ or, alternatively, \gls{ppp} of length = 2. 
The sub-areas in Figure~\ref{fig:voronoi}(b) are labeled ``$i,j$'' and they cover all objects $o \in \mathcal{D}$ for which ${\Pi^o_1}=i$ and ${\Pi^o_2}=j$. 
For example, data point $X$ induces permutation prefix ``$<6,4>$'' and data point $Y$ is ``$<3,2>$''.
\gls{ppp} avoids excessive fragmentation of the space using shorter representation while preserving the locality properties of {\em Pivot Permutation}. 

For similarity search queries, the distance between \gls{ppp}es is considered  
a good approximation for the distance between the original points. 
Thus, it should be possible to efficiently retrieve similar objects by examining only a tiny subset of data points whose permutation prefixes  
are similar to that of the query object.

\vspace{2mm}
{\textbf{Challenges in Adopting Pivot Permutation Prefix Techniques.}}
There are critical challenges in adopting the existing techniques  as is to
the data series domain. including: 
(1)~{\em Curse of Dimensionality}. Pivots are points in the original 
multi-dimensional space ($n$ dimensions per Def.~\ref{def:data-series}). 
If the space is very high dimensional, as in our case, then similarity measures between the data points and the pivots become meaningless and all points will appear to be too far from each other. 
(2)~{\em Index-Unfriendly Rigid Representation.} 
Existing techniques rely solely on ordered (rank-sensitive) vector 
representations, which is perfectly acceptable for 
direct in-memory comparisons on a centralized machine~\cite{amato2008approximate,gonzalez2008effective,chavez2005proximity, novak2016ppp,esuli2012use}. 
However, in our context, we 
aim for building a distributed disk-persistent 
index structure to support  search operations. 
These indexes are typically multi-layers and  require 
flexible granularities (i.e., coarse- and fine-grained representations) 
for efficient data partitioning. 
(3)~With the introduction of flexible coarse- and fine-grained representations, 
existing distance metrics in the literature, e.g.,Spearman's Footrule and Kendall's $\tau$~\cite{10.172749}, will not work, especially when comparing objects 
of different granularities. 
In the rest of this section and in Section~\ref{sec:idx}, we present methods for addressing these challenges within \gls{infra}.

%====================
\subsection{\gls{infra} Feature Extraction Strategies}
\label{sec:sign}

In this section, we present \gls{infra}-FX, the feature extraction component of our system. 
In a nutshell, \gls{infra}-FX addresses the first two challenges highlighted above by  (1)~reducing the data series dimensionality using segmentation techniques, e.g., PAA~\cite{keogh2001dimensionality}, 
and (2)~generating a novel dual representation that consists of 
{\em rank-sensitive} and {\em rank-insensitive} Pivot-Permutation-Prefix signatures, which are leveraged in a complementary fashion during the index construction process.

\begin{figure}[t]
  \centering
  \begin{subfigure}[t]{0.49\columnwidth}
      \centering
      \includegraphics[width=1.0\textwidth]{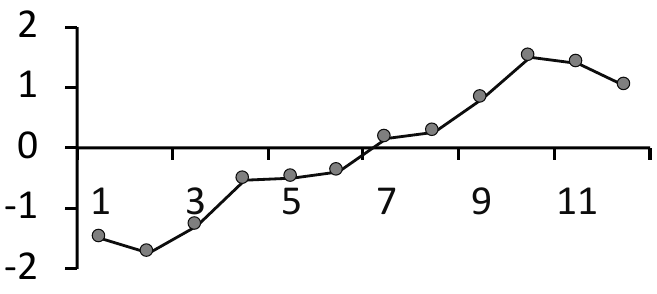}
      \caption{Data Series X, $n=12$}
      \label{fig:paa-1}
  \end{subfigure}%
  ~
  \begin{subfigure}[t]{0.49\columnwidth}
      \centering
      \includegraphics[width=1.0\textwidth]{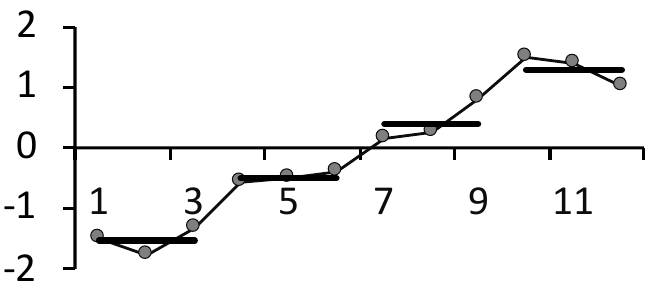}
      \caption{$PAA_X$=[-1.5,-0.4,0.3,1.5]}
      \label{fig:paa-2}
  \end{subfigure}%

  \caption{PAA segmentation of a data series.}
  \label{fig:paa}
  \vspace{-4mm}
\end{figure}

\vspace{1mm}
\textbf{Step 1. Segmentation of Data Series.} 
The raw data series objects may consist of 100s of readings. This  high dimensionality leads to the well-known ``curse  of dimensionality'' problem. 
Therefore, we first leverage a segmentation and encoding algorithm known as 
{\em Piecewise Aggregate Approximation} (PAA)~\cite{keogh2001dimensionality}. PAA 
is an equal-size segmentation strategy commonly used as a 
pre-processing step
due to its low complexity and robustness
e.g.,~\cite{camerra2010isax,kondylakis2019coconut,lin2007experiencing,shieh2008sax,yagoubi2017dpisax,zhang2019tardis,zoumpatianos2016ads}.

Given a raw data series $X$ of length $|X| = n$ 
and a pre-defined number of segments $w$, where $w \ll n$, 
PAA works as follows (refer to Figure~\ref{fig:paa}). 
First, $X$ is divided into $w$ segments, and then for each segment its mean value is calculated. The resulting vector (referred to as $PAA_X$) is the 
PAA signature of $X$ in the lower dimensional space $w$, e.g., 
in the example illustrated in Figure~\ref{fig:paa} the dimensionality is reduced from 
$n = 12$ to $w=4$. 
PAA is a lossy technique, but it is not detrimental 
as long as the segment length $\frac{n}{w}$ is  kept small.
Although the mean value calculation of each segment is lossy, 
PAA signatures preserve the similarity better than iSAX signatures
because the PAA similarity is evaluated based on the mean values rather than the stripe labels as in iSAX. For example, referring to Figure~\ref{fig:sax-isax-rep}, 
under PAA, the distances between ($a$, $b$) and ($c$, $d$) are now different and 
more accurate compared to the iSAX signature.

\vspace{1mm}
\textbf{Step 2. $P^4$ Dual Signature Generation.} 
In this step, we apply \textit{Pivot Permutation Prefix} (Def.~\ref{def:ppp}) 
on the PAA signatures to generate our dual representations, which we refer to as $P^4$.

\begin{figure}[t]
  \centering
  \includegraphics[width=1.0\columnwidth]{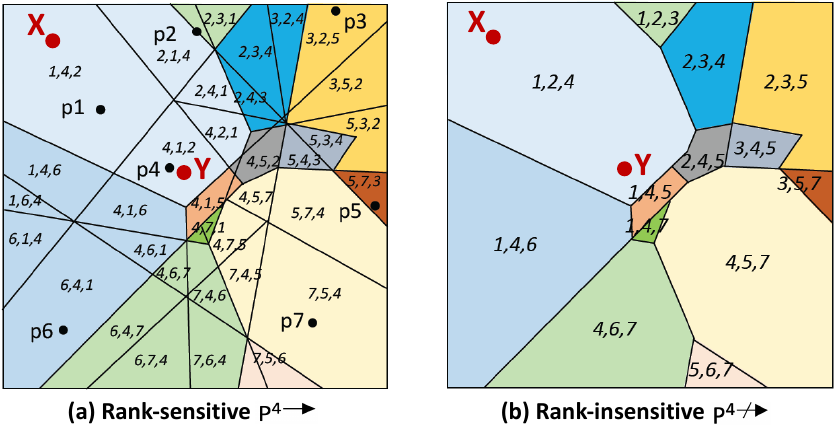}
  \caption{$p^4$ Signatures where $\#$ of prefix pivots $m=3$.}
  \label{fig:pivot}
  \vspace{-3mm}
\end{figure}

\begin{definition}
\label{def-dual}
  \textbf{[$\bm{P^4}$  Dual Signature]} 
  \textit{
      Given a PAA representation of data series $X$ ($PAA_X$), a set of $r$ pivots 
      $\{p_1, p_2, \cdots, p_r\}$, and a prefix length $m$,  the rank-sensitive 
      and rank-insensitive signatures of $X$ are denoted as $P^{4\rightarrow}_X$ and $P^{4\nrightarrow}_X$, respectively, where:\\
       $~~~~~~~$ $P^{4\rightarrow}_X$ = $PPP(PAA_X)$ as per  Def.~\ref{def:ppp}, and\\ 
      $~~~~~~~~P^{4\nrightarrow}_X$ = LexicographicalOrder( $P^{4\rightarrow}_X$)
  }
\end{definition}

In other words, \gls{p4i} signature ignores the distance ranking of pivots and orders the pivots by their identifiers (Ids). 
In Figure~\ref{fig:pivot}, we illustrate the idea of the rank-sensitive and rank-insensitive signatures. 
For example, for the data series objects $X$ and $Y$ in Figure~\ref{fig:pivot}, the $P^{4\rightarrow}_X$ and $P^{4\rightarrow}_Y$ signatures 
correspond to ``$<1, 4, 2>$'' and  ``$<4, 1, 2>$'', respectively 
(Figure~\ref{fig:pivot}(a)). Whereas, both objects share the same rank-insensitive representation of $P^{4\nrightarrow}_X = P^{4\nrightarrow}_Y = $ ``$<1, 2, 4>$'' (Figure~\ref{fig:pivot}(b)). 
    As depicted in Figure~\ref{fig:pivot}, the two representations 
    impose two levels of granularities for data partitioning. 
    That is, the rank-insensitive representation creates coarse-grained 
    partitioning while the rank-sensitive representation 
    achieves fine-grained partitioning. 
   We propose  these two levels of granularities as critical
    components of our
 hierarchical index design.
    More specifically, the two representations will be leveraged to  
    cluster similar data series objects together into   
    a two-level hierarchy structure; namely, {\em Data Series Groups} and 
    {\em Data Series Partitions}. The $1^{st}$ level relies on the 
    rank-insensitive \gls{p4i} signatures to form the groups 
    (Section~\ref{sec:group_dist}), while the $2^{nd}$ level relies on 
    the rank-insensitive \gls{p4s} signatures to form the partitions (Section~\ref{sec:partitions}).

\subsection{Data Series Group Formation}
\label{sec:group_dist} 

Data series groups are coarse-grained clusters created 
in the rank-insensitive feature space. Since existing techniques in the literature, e.g.,~\cite{amato2008approximate,gonzalez2008effective,chavez2005proximity}, operate only on the rank-sensitive signatures, 
they use distance functions such as Spearman Footrule and Kendall's $\tau$ distances~\cite{10.172749}. 
However, our group formation process involves rank-insensitive comparisons. 
Therefore, we only care about the number of pivot mismatches to measure the distances, and thus  use the overlap distance function defined as follows:

  \vspace{1mm}
\begin{definition}
\label{def.overlap}
  \textbf{[Overlap Distance]} 
  \textit{Given two rank-insensitive signatures $P^{4\nrightarrow}_X$ and $P^{4\nrightarrow}_Y$ for data series $X$ and $Y$, where $|P^{4\nrightarrow}_X| =|P^{4\nrightarrow}_Y| =m$, the Overlap Distance is defined as:}
  \begin{equation}
    OD(X,Y) = m -  P^{4\nrightarrow}_X \cap P^{4\nrightarrow}_Y \label{def:od}
  \end{equation}
  $P^{4\nrightarrow}_X \cap P^{4\nrightarrow}_Y \geqslant 0$, \textit{ thus, $OD(X,Y)$ lies in the range $[0,m]$}.
  \vspace{-1mm}
\end{definition}

In other words, the overlap distance between two \gls{p4i} signatures is the prefix length $m$ minus the intersection's cardinality of two sets. 
For example, assume $P^{4\nrightarrow}_X = <1, 3, 6, 8>$ and $P^{4\nrightarrow}_Y = <2, 3, 4, 6>$, 
then $OD(X,Y) = 4 - 2 = 2$.

\vspace{1mm}
A data series group is then defined as follows:

\begin{definition}\label{def:group}
  \textbf{[Data Series Group]} 
  \textit{
    Given a data series dataset $DB$ and a set of size $k$ of pre-selected centroids, each is in its 
    rank-insensitive representations $P^{4\nrightarrow}_{oi}~\forall 1\leq i \leq k$, 
    a Data Series Group $G_i$ having centroid $P^{4\nrightarrow}_{oi}$ is define as  
    $G_i = \{ X_1, X_2, \cdots, X_m\} \big| \forall X_l \in G_i, 
    OD(P^{4\nrightarrow}_{oi},P^{4\nrightarrow}_{X_l}) \leqslant 
    OD(P^{4\nrightarrow}_{oj},P^{4\nrightarrow}_{X_j}), i \neq j$. 
  }
\end{definition}

For simplicity, we assume for now that the $k$ centroids are pre-computed
along with their rank-insensitive representations. 
It is important, however, to mention that these centroids are virtual points, 
and hence unlike the real data series objects, these centroids 
do not have a  rank-sensitive representation.
Later in Section~\ref{sec:idx}, we present an efficient algorithm for generating these centroids based on the data distribution.

For the groups' formation, each data series object, say $X$, is assigned to the closest centroid based on the distance function in Def.~\ref{def.overlap}. 
However, it is possible that ties may exist if more than one group shares the same smallest $OD$ distance. 
In this case, instead of randomly selecting one of these groups, we 
leverage the fact that $X$ has a more detailed 
rank-sensitive representation. Therefore, 
we propose a secondary distance function 
that learns, based on $X$'s rank-sensitive signature, the pivots' importance to $X$ and assigns them weights accordingly. These weights are used to refine the distance calculation and break a tie 
as follows.

\begin{definition}\label{def:weight}
  \textbf{[Pivots Weight]} 
  \textit{Given a data series object $X$ and its rank-sensitive signature 
  $P^{4\rightarrow}_X = <D_{\Pi^X_1}, D_{\Pi^X_2}, \cdots, D_{\Pi^X_m}>$, the weight of each pivot for $X$ is derived based on a decay function f() and a decay rate $\lambda \in (0,1)$ 
  as  $W(D_{\Pi^X_i}) = f(i, \lambda)$ such that:}
\begin{center}
  $W(D_{\Pi^X_1}) > W(D_{\Pi^X_2}) > ... > W(D_{\Pi^X_m})$
\end{center}
\end{definition}

Def.~\ref{def:weight} indicates that in $P^{4\rightarrow}_X$ the pivots appearing from left-to-right are more important and hence should receive higher weights compared to the succeeding pivots in a signature.
This is because the $1^{st}$ pivot from left is the closest to $X$, and 
then the $2^{nd}$ pivot from left is the  $2^{nd}$ closest to $X$, and so on.
Examples of decay functions include the exponential decay, e.g., $f(i, \lambda)=\lambda^{(i - 1)}$, 
and linear decay, e.g., $f(i, \lambda)= \lambda * (m - i + 1), \lambda = \frac{1}{m}$. 
For example, if $\lambda=1/2$, then the exponential decay sequence is [1, 1/2, 1/4, ...] 
while the linear decay sequence is [1, m-1/m, m-2/m, ...], where $m$ is the length of the pivot prefix.

The total weight for a given \gls{p4s} signature can be computed as described below.

\begin{definition}
  \textbf{[Total Weight]} 
  \textit{Given the \gls{p4s} signature of a data series X,  $P^{4\rightarrow}_X = [D_{\Pi^X_1}, D_{\Pi^X_2}, \cdots, D_{\Pi^X_m}]$, the total weight of $P^{4\rightarrow}_X$ is defined as:}
  \begin{equation}
    TW(X) = \sum_{i=1}^{|P^{4\rightarrow}_X|} W(D_{\Pi^X_i}) 
  \end{equation}
  \vspace{-2mm}    
\end{definition}

Note that given the number of pivots for all data series objects is the same, and the decay function is fixed,  the {\em Total Weight} for all \gls{p4s} signatures becomes a constant value.

The secondary distance function that is used to break ties in the group assignment is next defined.

\begin{definition}
  \textbf{[Weight Distance]} 
  \textit{Given a rank-sensitive signature $P^{4\rightarrow}_X$  for a data series object
  and a rank-insensitive signature $P^{4\nrightarrow}_{o_i}$ for a group's centroid, the Weight Distance between them is defined as:}
  \begin{equation}
    \begin{aligned}
       WD(X, o_i) = & TW(X) - \sum_{i = 1}^{|P^{4\rightarrow}_X|} W(D_{\Pi^X_i}) * \mathbbm{1}_{D_{\Pi^X_i}} \\  
       \mathbbm{1}_{D_{\Pi^X_i}} = &
        \begin{cases}
        1\hspace{0.5cm} \text{if } D_{\Pi^X_i}\in P^{4\nrightarrow}_{o_i}\\
        0\hspace{0.5cm} \text{if } D_{\Pi^X_i}\notin P^{4\nrightarrow}_{o_i}
        \end{cases} 
    \end{aligned}
  \end{equation}
%  $\textit${WD(X, Y) lies in the range [0, TW(X)].}
\end{definition}

\begin{algorithm}[t]
\caption{Group Assignment Rules}
\label{alg:groupAssign}
\small{
      \KwIn{ List of centroids $S_c = \{o_1, o_2, ...\}$    // each has its $P^{4\nrightarrow}_{oi}$ sig., 
             special fall-back centroid $o_0$, 
             object $X$ with its $P^{4\rightarrow}_X$ and $P^{4\nrightarrow}_X$ sig.}
      \KwOut{centroid id //Assign $X$ to one of the centroids }
      \Begin{
            Calculate the OD distances OD($P^{4\nrightarrow}_{oi}$, $P^{4\nrightarrow}_X$) $\forall oi \in S_c$ \\
            \If{all OD distances = $m$ (// m is the sig. length)}
                { // no overlap with any pivot \\
                \textbf{return} $0$     // assign X to the fall-back centroid $o_0$} 
            \If{unique smallest OD distance (at position $v$)}
                {\textbf{return} $v$     // assign X to centroid $o_v$} 
            //A tie scenario occurs (say for subset of centroids $S'_c$) \\
            Calculate the pivot weights for $X$ per Def.~\ref{def:weight}   \\
            Calculate the WD distances WD($P^{4\rightarrow}_X$, $P^{4\nrightarrow}_{oi}$) $\forall oi \in S'_c$ \\
            \If{unique smallest WD distance (at position $v$)}
                {\textbf{return} $v$     // assign X to centroid $o_v$}
            // A $2^{nd}$ tie scenario occurs \\
            \textbf{return} (randomly selected centroid id from $S'_c$)
        }
    }
\end{algorithm}

The WD distance  corresponds simply to the total weight score (constant value) minus the sum of weights 
of $X$'s pivots present in the group's centroid $P^{4\nrightarrow}_{o_i}$. 
Therefore, the more of $X$'s pivots present in $o_i$ and the higher their weights, the lower the WD distance. 

The {\em Overlap} and  {\em Weight} distances can be leveraged for the group assignment as depicted in 
Algorithm~\ref{alg:groupAssign}. Simply put, if  a centroid  $o_v$ exists with a unique smallest OD distance, then $o_v$ is selected (Lines 2 \& 6-7). Otherwise, the pivot weights need to be calculated (Line 9) 
and then the WD distances are computed with the tie-involved centroids (Line 10). 
The centroid selected among  them is either the one with the unique smallest WD distance (if exists) (Lines 11-12), 
or a randomly selected centroid (Line 14). 
Lines 3-5 address  a special case when the data series object $X$ has zero overlap with any of  centroids. 
In this case, $X$ is assigned to a special fall-back centroid $O_0$.

\begin{example}
  \textit{
   Assume two pivot groups, namely $G_1$ and $G_2$, 
   having centroids $P^{4\nrightarrow}_{o1} = <1,2,3>$
   and $P^{4\nrightarrow}_{o2} = <2,4,5>$, respectively 
   (recall that centroids have only rank-insensitive signatures). 
   Assume also three data series object $X, Y, Z$ as 
   $P^{4\rightarrow}_X=<3,4,1>$
   $P^{4\rightarrow}_Y=<4,2,1>$
   and $P^{4\rightarrow}_Z=<6,2,7>$. 
   The objects' rank-insensitive signatures are 
   $P^{4\nrightarrow}_X = <1,3,4>$
   $P^{4\nrightarrow}_Y = <1,2,4>$
   and $P^{4\nrightarrow}_Z = <2,6,7>$.
   In this example, we use exponential decay function for tie breaks.
  The assignment is then performed as follows:
  \begin{center}
  {\small{
  \begin{tabular}{ l|l } 
  \hline
  $OD(X,G_1.o_1)$=1  & Assign X to $G_1$  \\ 
  $OD(X,G_2.o_2)$=2  &   \\\hline
  $OD(Y,G_1.o_1)$=1  & Calculate the weights of Y's pivots \\
  $OD(Y,G_2.o_2)$=1  & W(4)=1.0, W(2)=0.5,W(1)=0.25 \\
    (tie)           & Total 2eight = 1.75  \\
                 & WD(Y, $G_1.o_1$) = \\
                 &1.75 – (W(1) + W(2)) = 1 \\
                 & WD(Y, $G_2.o_2$) = \\
                 & 1.75 – (W(4) + W(2)) = 0.25 \\
                 & Assign Y to $G_2$\\\hline
 $OD(Z,G_1.o_1)$=2  & Calculate weights of Z's pivots \\
 $OD(Z,G_2.o_2)$ = 2 & W(6) = 1.0, W(2) = 0.5, W(7) = 0.25  \\
    (tie)         & Total weight = 1.75 \\
                & WD (Z, $G_1.o_1$) = 1.75 – (W(2)) = 1.25 \\
                & WD(Z, $G_2.o_2$) = 1.75 – (W(2)) = 1.25 \\
                & Assign Z randomly to either $G_1$ or $G_2$\\
 \hline
\end{tabular}
}}
\end{center}
  }
\end{example}

\subsection{Data Series Partition Formation}
\label{sec:partitions} 

Since there is no size constraints on the {\em Data Series Groups}, their sizes can be large. 
Therefore, for efficient indexing and retrieval, we 
propose a group-level partitioning strategy to divide such large groups 
into {\em Data Series Partitions}. As will be presented next, 
this partitioning strategy divides the space into Voronoi-based partitions 
based on pivot permutations. Hence, the partitions are
aligned with the pivot-based space fragmentation 
(see the partitioning from Figure~\ref{fig:pivot}(b) to Figure~\ref{fig:pivot}(a)). 
These partitions will serve as 
the physical storage units in CLIMBER.

\vspace{1mm}
\begin{definition}\label{def:partition}
  \textbf{[Data Series Partition]} 
  \textit{
    Given a data series group~$G$ and a capacity constraint $c$, 
    the goal is to partition $G$ into a set of Voronoi-based partitions 
    \{$\beta_1, \beta_2, ..., \beta_m$\} such that: 
    (1)~partitions are disjoint, 
    (2)~partitions achieve full coverage, i.e.,  $\bigcup^{m}_{i=1} \beta_i = G$, and 
    (3)~capacity of each partition is less than or equal to $c$. 
    If the capacity of~$G$ is less than or equal to $c$, then
    $G$ is considered as a single partition.
  }
\end{definition}

\begin{figure}[t]
  \centering
  \includegraphics[width=0.95\columnwidth]{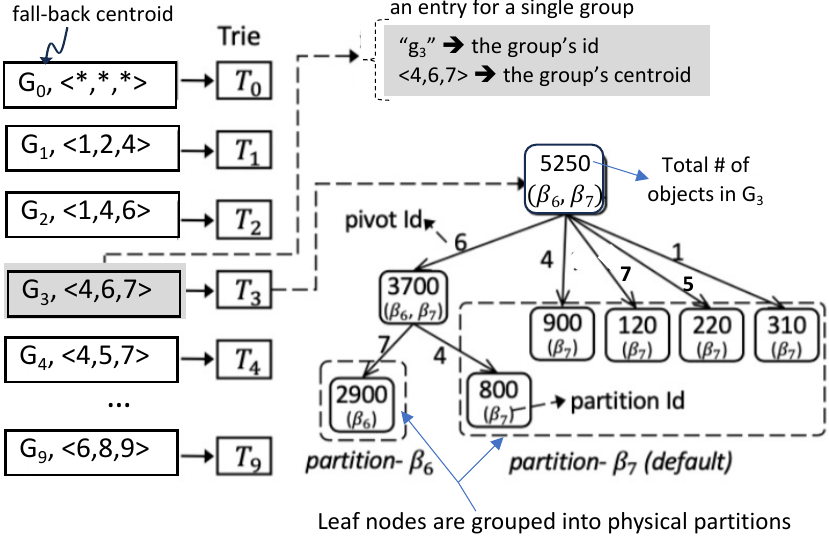}
  \caption{\gls{infra}-INX index skeleton: groups (the $1^{st}$ Level) and trie-based partitions (the $2^{nd}$ level).}
  \label{fig:trie}
  \vspace{-4mm}    
\end{figure}

The capacity constraint $c$ is mainly determined by the underlying storage system, e.g., 
for Apache Hadoop Distributed File System (HDFS)~\cite{shvachko2010hadoop}. Typically, $c$ is $64$ \textit{MB} or $128$ \textit{MB}.

To construct the Voronoi-based partitions for a given group, 
we propose a trie-based splitting strategy that works as follows. 
Assume the capacity constraint c=3,000. Then, for each group $G$ with higher capacity than $c$, we 
distribute $G's$ data series objects based on their $1^{st}$ pivot 
in their \gls{p4s} signatures to form the $1^{st}$ level of the trie  structure associated with $G$. 
For example, Group $[G_3,<4,6,7>]$ in  Figure~\ref{fig:trie} 
consists of a total of 5,250 objects (the total count is present in the root node), 
and hence its objects  are distributed based on their 
distinct $1^{st}$ pivots as illustrated in the figure. 
For example, pivot ``6'' receives 3,700 objects, pivot ``4'' receives 900 objects, and so on. 
Since the partition under pivot ``6'' is still greater than $c$, the node recursively 
splits its objects based on the  distinct $2^{nd}$ level pivots within this group. 
This process continues until all leaf nodes in the trie have a capacity less than $c$. 

The trie-based partitioning has several desirable properties including: 
(1)~it creates Voronoi-based partitions where each leaf node has its root-to-leaf path representing its pivot prefix, 
(2)~it is simple and light-weight, 
and (3)~it provides flexibility for packing multiple leaf nodes together 
in a single partition to avoid having many tiny partitions--with the later prohibitive for distributed systems.

Notice that when creating a group's trie, some nodes in the trie may have pivot ids that do not appear 
in the group's signature. 
For example, the trie associated with Group $[G_3,<4,6,7>]$ has pivots such as ``5'' and ``1'', which is acceptable because some objects within the group may have signatures not exactly matching the group's centroid signature. 
After creating a group's trie, the  leaf nodes are packed into partitions, 
and each node--leaf or internal--is labeled with its corresponding 
partition id.
For example, five leaf nodes in Figure~\ref{fig:trie} are packed into one partition $\beta_7$, while  the left-most leaf node is packed 
into a separate partition $\beta_6$. The internal nodes are labeled with the partition ids representing the union of their subtrees.
The packing algorithm is presented in detail in Section~\ref{sec:idx}. 

Group $G_0$ is reserved for data series that can't find proper groups.  Specifically, the \gls{p4i} signature of the group centroid don't contain any pivot in the \gls{p4i} of data series, i.e., the {\em Overlap Distances} between the data series and all centroids equal to $m$.

%===================================================
%===================================================

\section{Indexing Framework}
\label{sec:idx}

\begin{figure}[t]
  \centering
  \includegraphics[width=0.99\columnwidth]{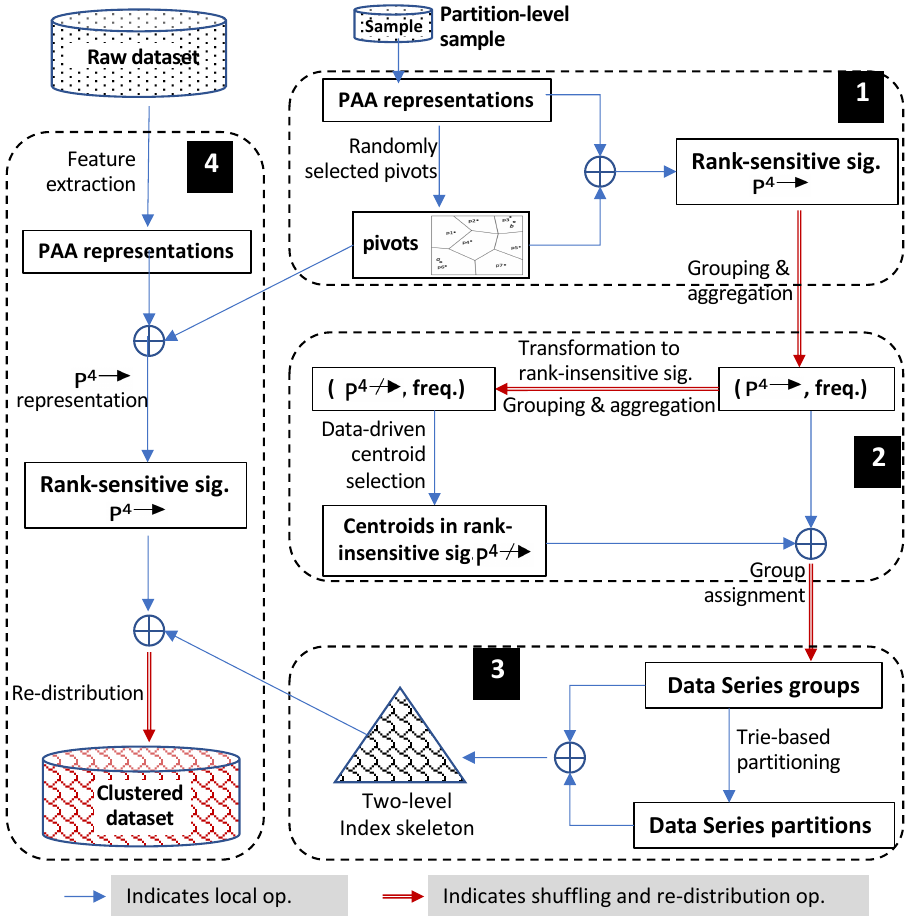}
  \caption{Index construction workflow.}
  \label{fig:data-flow}
  \vspace{-2mm}    
\end{figure}

In this section, we present the details of the indexing framework, namely CLIMBER-INX. 
The construction process is illustrated in Figure~\ref{fig:data-flow} and it 
consists of four main steps. The first three steps rely solely on a small sample from the data to ultimately generate the index skeleton (the partitioning scheme). 
The sample is generated at the partition level, i.e., 
a subset of the data partitions are randomly selected.
This way full-scan over the data is avoided. 
We assume that this sample is representative because the original dataset  
in most applications gets stored across partitions without any special or custom organization.
Finally, the fourth step accesses and re-distributes the entire dataset to generate 
the final data partitions based on the index skeleton.

\textbf{Step 1. Generation of Rank-Sensitive Signatures.} 
In this step, each data series in the sample is converted 
into its PAA signature based on a configurable 
segment length parameter $w$.
Then, a subset of size $r$ is randomly selected to form the set of pivots  $\{p_1, p_2, \cdots, p_r\}$. 
We opt for random selection because  
existing work in literature~\cite{novak2011metric,esuli2012use,gonzalez2008effective,tellez2011succinct,novak2016ppp} has shown that random selection works competitively well compared to any other sophisticated selection methods. 
These pivots remain fixed throughout the entire system operations. 
Finally, all PAA signatures are converted to their \gls{p4s} signatures 
per Def.~\ref{def-dual}.

\textbf{Step 2. Computation of Group Centroids.} 
The goal of Step 2 is to devise a data-driven method for 
computing the centroids based on which the data series groups will be formed. 
Recall from Section~\ref{sec:group_dist} that these groups  
represent  coarse-grained clusters created in the rank-
insensitive feature space. These groups serve as 
the index's $1^{st}$ level hierarchy. 

Algorithm~\ref{alg:part} sketches the procedure for computing the groups' centroids.
The key intuition is to create groups (clusters) that have high membership and also provide good coverage for the entire space, i.e., centroids are relatively far from each other. 
We achieve these goals 
as follows. The output from Step 1, which is a list of $P^{4\rightarrow}$ signatures,  
is aggregated based on exact matching to create a list of pairs [($P^{4\rightarrow}$, frequency)]. 
Then, each $P^{4\rightarrow}$ signature is converted to it's corresponding $P^{4\nrightarrow}$ and 
also aggregated based on exact matching to create a list of pairs [($P^{4\nrightarrow}$, frequency)] 
(see Step 2 in Figure~\ref{fig:data-flow}). 
This list becomes List L that is passed as input to Algorithm~\ref{alg:part}. 

In Line 1 in Algorithm~\ref{alg:part}, we sort L descendingly and then pick L[0] (the signature with the highest frequency) as the $1^{st}$ centroid (Line 3). 
Then, we move down in the list, one at a time, and add a given L[i] signature to the centroids' list if 
it satisfies two conditions: (1) L[i] is far enough from all previously selected centroids (Lines 5-9). 
This is to avoid selecting very close centroids and to have a good coverage in the space. 
And (2) the group associated with centroid L[i] is expected to have number of members larger than 
the capacity threshold $c$ (Lines 10-13). 
This size can be estimated using the equation in Line 11 (assuming a uniform distribution of the remaining signatures over the current centroids). 
%We know the frequency associated 
%with L[i]--these are signatures that are guaranteed to belong to this new group 
%(if L[i] is selected as a centriod). In addition, all non-centroid signatures in L will get distributed across the different groups. Assuming a uniform distribution, we can estimate the additional signatures that might join the group centered by L[i] as depicted in Line 9.  
Notice that in Line 12, $c$ is multiplied by $\alpha$ (the sample percentage) since the statistics computed in Step 2 are based on a sample instead of the entire datasets.
The output from Algorithm~\ref{alg:part} 
is a list of centroids and each one is assigned a unique group id 
$[ (group\text{-}id, P^{4\nrightarrow}), ... \big]$ 
(see the left side of Figure~\ref{fig:trie}).

\begin{algorithm}[t]
      \caption{Computation of Groups' Centroids}
      \label{alg:part}
\small{      
      \KwIn{ $\alpha$ //sample $\%$, $c$ //storage capacity constraint, 
             $L = [(P^{4\nrightarrow}, freq)]$ //rank-insensitive signatures and their freqs., $MaxCentroids$ //optional stopping criterion }
      \KwOut{List of centroids $S_c$, initially $S_c = []$}
      \Begin{
          $L~\leftarrow$ Sort list $L$ descending based on $freq$ values;\\
          Add L[0] to $S_c$;  //highest freq. $P^{4\nrightarrow}$ is the $1^{st}$ centroid.\\
          \For{(i = 1, i++, i $<$ Len(L)) }{
            \For{(j = 0, j++, j $<$ Len($S_c$)) } { 
               //check dist. against existing centroids\\
               //(avoid too close centroids)\\
              \If{OverlapDist(L[i].sig, $S_c$[j].sig) $<~\epsilon$ }{
                 Break the loop and go for the next i from L;   
                }
              }
              //avoid tiny groups\\
              sizeEst(L[i]) = L[i].freq + (sum freq. of non-centroids / (Len($S_c$)+1)); \\
              \If{sizeEst(L[i]) $<$ ($\alpha \times c$)} {
                 Set $S_c$ is final, break the loop;  
              }
              Add L[i] to $S_c$;  //one more centriod is identified \\
              \If{Len($S_c$) == MaxCentroids}{
                 Set $S_c$ is final, break the loop;               
              }
              
          }
          Add special centroid $<*$,$*$,...$>$ to $S_c$;\\
          \textbf{return} \textit{$S_c$}
    }
}
\end{algorithm}

\textbf{Step 3. Construction of Groups and Partitions.}
After generating the groups' centroids, the next step is to form the groups
by distributing the rank-sensitive signatures (list [($P^{4\rightarrow}$, freq.), ...], which is coming from Step 2  
to their nearest centriods (see Figure~\ref{fig:data-flow}). 
This assignment is performed for each $P^{4\rightarrow}$ based on the rules described in  Algorithm~\ref{alg:groupAssign}.

Once the groups are formed, the size of each group can be estimated, i.e., the summation of all frequencies from the $P^{4\rightarrow}$ members, and then multiplied by 100/$\alpha$, where $\alpha$ is the sampling percentage. 
If a group size is expected to be larger than the capacity threshold ($c$), 
then the trie-based partitioning 
kicks in to create the proper trie nodes of that group as described in Section~\ref{sec:partitions} and Figure~\ref{fig:trie}.
Note that the root-to-leaf paths with a given trie is labeled with the pivots 
prefixes leading to this leaf node. Moreover, the size of each node can be estimated 
in a similar way to that of group size estimation. 

The last step is to pack the leaf trie nodes into as few partitions as possible (these will be the physical storage partitions) such that a partition size does not exceed a storage capacity $c$. This problem can be formulated as follows. 

\vspace{1mm}
\begin{definition}\label{def:leafpacking}
  \textbf{[Node Packing Problem]}
    \textit{
      Given a list of leaf nodes $N = \lbrace n_1,n_2,\cdots,n_m\rbrace$ and a partition capacity $c$, the node packing problem is to group leaf nodes into as few partitions as possible, such that the sum for each partition is no greater than $c$.
    }
  \end{definition}
  
It can be shown that this packing problem is a well-known 
NP-hard optimization problem \cite{delorme2016bin}.
Therefore, we adopt the \emph{First Fit Decreasing} (FFD) algorithm presented in ~\cite{delorme2016bin}, which is the best known approximation algorithm for this problem with a time complexity of $O(m~log(m))$ and worst-case performance ratio $1.5$ from the optimal solution. 
The algorithm is a greedy sort-based technique, but we omit the details due to space limitations. 

The construction process of the index skeleton completes 
by assigning a unique id for each  physical partition 
and linking these ids to the trie nodes (leaf and internal) 
as illustrated in Figure~\ref{fig:trie} by ids $\beta_6$ and $\beta_7$.
Since the index skeleton is constructed based on a sample from the input dataset, it is worth mentioning two points: (1)~The storage capacity constraint $c$ is typically a soft constraint, and thus when processing the entire dataset, the final partition sizes could slightly differ, and (2)~It is possible that when processing the entire dataset  we counter signatures that have not been seen in the sample. 

The handling of the unseen data series signatures is addressed at the group level by having the special fall-back group $G_0$ (as described in Section~\ref{sec:group_dist} and Algorithm~\ref{alg:groupAssign}). 
Moreover, it is addressed  at the partition level by assigning for each group 
a default physical partition$-$typically the partition with the smallest occupancy. 
For example, group $G_3$ in Figure~\ref{fig:trie} has $\beta_7$ as the default partition 
since it has less data compared to $\beta_6$. Any data series member of 
group $G_3$ that cannot navigate a complete root-to-leaf path in the trie goes to 
the group's default partition.

\textbf{Step 4. Data Re-Distribution based on dual pivot-based similarities.}  
This step involves the distributed processing of the entire dataset and re-distributing the 
data series objects into the appropriate groups and partitions. 
For that, both the set of pivots (from Step 1) and the index skeleton (from Step 3) are broadcasted to all machines in the cluster. 
The size of the pivots and the index skeleton is tiny and 
they can fit easily the the memory of each machine.
The set of pivots are needed for the generation of the  $P^{4\rightarrow}$ and $P^{4\nrightarrow}$ 
signatures for each data series object, while the index skeleton is needed for identifying the group and partition assignments (see Figure~\ref{fig:data-flow}). 

The final output from Step 4 represents a set of physical partitions that 
cluster the original data series objects according to their dual pivot-based   
similarities. In short, the groups impose rank-insensitive similarities while partitions 
impose rank-sensitive similarities.

%============================================================
%============================================================

\section{CLIMBER-kNN Query Processing Strategies}
\label{sec:query}

{\textbf{CLIMBER-kNN Algorithm.}}
In this work, we target  kNN approximate similarity queries
as defined in Def.~\ref{def:knnapprox}
as core query operation 
~\cite{zhang2019tardis,echihabi2019return,yagoubi2017dpisax,nourachainlink,wu2019kv,zhang2020big,isaxfamily,arora2018hd}. 
Given a query data series $X_q$, the 
 procedure for answering the query is presented in Algorithm~\ref{alg:kNN}. 
First, $X_q$ follows the same transformation steps that each data series follows 
while creating the index (see Step 4 in Figure~\ref{fig:data-flow}). 
The PAA representation of $X_q$ is first generated ($PAA_{X_q}$).
Then,  given the set of 
pre-selected pivots in the system, 
the rank-sensitive and rank-insensitive representations of $X_q$ are 
derived, i.e., $P^{4\rightarrow}_{X_q}$ and 
$P^{4\nrightarrow}_{X_q}$, respectively (Lines 2-4). 
Second, the $P^{4\nrightarrow}_{X_q}$ signature is used to navigate the CLIMBER-INX index skeleton 
to find the best matching data series group(s) based on the OD distance (Lines 5-6). 
If there are multiple such groups sharing the same best score, 
the algorithm uses the $P^{4\rightarrow}_{X_q}$ signature to compute the pivots' 
    weights for $X_q$ and then calculates the WD distances to break the ties and 
    select the best matching group (Lines 7-9). 
    It is  possible to have a second tie, which means 
 multiple candidate groups may be considered.

The next step is that for each candidate group  $G$, 
    the $G$'s trie is traversed top-down 
    based on the $P^{4\rightarrow}_{X_q}$ signature 
    until the navigation path cannot be extended further.
    For instance, we may 
   stop at trie node $G_N$, which can be a root, an internal, or a leaf  (Lines 10-13). 
    In the case, we have more than one candidate group,  
    we  break the tie by 
    selecting the group having the longest matching path to its $G_N$, 
    and then the one having the largest node size (Lines 14-17). 
    Recall that the deeper we traverse in a trie, the better the matching to the query object becomes. Moreover, the larger the node size is the more candidate data series objects to examine. 
    In the rare cases where a tie is remains,  we  apply a random selection  to select one among these already well matching groups (Lines 18-19).
    Finally, give an identified group $G$ and its best matching trie node $G_N$, the physical partitions associated with $G_N$ are added to the output list.

\begin{algorithm}[t]
      \caption{CLIMBER-kNN Algorithm}
      \label{alg:kNN}
\footnotesize{      
      \KwIn{ Query object $X_q$}
      \KwOut{$ParOutput = []$  // List of physical partition ids}
      \Begin{
          $PAA_{X_q}$ $\leftarrow$ compute the PAA representation of $X_q$;\\
          $|P^{4\rightarrow}_{X_q}|$ $\leftarrow$ compute the rank-sensitive sig. of $X_q$; \\
          $|P^{4\nrightarrow}_{X_q}|$ $\leftarrow$ compute the rank-insensitive sig. of $X_q$; \\

          Calculate the OD distance OD(g, $|P^{4\nrightarrow}_{X_q}|$) $\forall~g$ in CLIMBER-INX;\\
          GList $\leftarrow$ the group(s) with the smallest OD distance;\\
          \If{Len(GList) $>$ 1 }{
            Calculate the WD distance WD($|P^{4\rightarrow}_{X_q}|$, g) $\forall~g$ in GList;\\
            GList $\leftarrow$ the group(s) with the smallest WD distance;\\
          }

          \For{each group G in GList}{
            Traverse G's trie top-down based 
            on the $|P^{4\rightarrow}_{X_q}|$ signature 
            to reach the deepest possible node (Node $G_N$). \\
            Let Size($G_N$) be the number of data series objects in $G_N$;\\
            Let PathLen($G_N$) be the path length from the root to $G_N$;\\
          }

          \If{Len(GList) $>$ 1 }{
            GList $\leftarrow$ the group(s) whose PathLen($G_N$) is the longest;\\            
          }
          
          \If{Len(GList) $>$ 1 }{
            GList $\leftarrow$ the group(s) whose Size($G_N$) is the largest;\\            
          }

           \If{Len(GList) $>$ 1 }{
            GList $\leftarrow$ select a single group randomly from GList; \\            
          }
          //GList contains a single group (say $G$)\\
          ParOutput $\leftarrow$ the physical partition(s) associated with $G_N$;
          
          \textbf{return} \textit{ParOutput}
    }
    }
\end{algorithm}

\begin{example}
{\em{Assume a query data series $X_q$ having signatures 
    $P^{4\rightarrow}_{X_q} = <6,2,7>$ and $P^{4\nrightarrow}_{X_q} = <2,6,7>$
    and the index skeleton presented in Figure~\ref{fig:trie}. 
    The CLIMBER-kNN algorithm selects Group $G_3$ as the best matching group 
    based on the OD distance, i.e., $OD(P^{4\nrightarrow}_{X_q}, G_3) = 1$. Then, the navigation inside 
    $G_3$'s trie based on $P^{4\rightarrow}_{X_q} = <6,2,7>$ results in stopping 
    at the internal node having capacity 3700. As a result, the algorithm returns the physical partitions association with this node (i.e., $\beta_6$ and $\beta_7$) as the output partitions.  
}}
\end{example}

{\textbf{CLIMBER-kNN-Adaptive Algorithm.}}
    This algorithm is a simple  variation of Algorithm~\ref{alg:kNN} yet effective 
    in boosting the query accuracy 
    in certain  scenarios. Namely,
    it targets the  case  when the best matching trie 
    node from Algorithm~\ref{alg:kNN} (Node $G_N$) contains less than the required output size $k$. 
    Although the physical partition(s) associated with $G_N$ still contain  more data series objects 
    than $k$, the query's accuracy may  degrade because the other trie nodes packed into the same partition(s)
    may not be as close to the query object $X_q$ as Node $G_N$. 

    To handle this   scenario, the CLIMBER-kNN-Adaptive algorithm memorizes more best matching groups, e.g., all groups having the same smallest OD distance or having a distance less than a certain threshold. 
    Within these groups, it also memorizes the best matching trie nodes, e.g., the longest and $2^{nd}$ longest best matches in each group. 
    If this scenario  is encountered, the algorithm 
    automatically expands its selection to multiple best 
    matching trie nodes such that the sum of their sizes is larger than $k$. 
    Then, the physical partitions associated with these trie nodes are 
    added to the output list of the algorithm. 
    CLIMBER-kNN-Adaptive can also leverage a configuration parameter 
    ({\em MaxNumPartitions}) for upper bounding the number of accessed partitions. 

\vspace{1mm}
{\textbf{Localized Record-Level Similarity within Identified Partitions.}}
Given an identified data partition (or a set of partitions) returned
by the above  search algorithms along with a specific trie node(s) to target, CLIMBER  finalizes the query answer  as follows. 
The data records within each data partition are organized such that all data series objects belonging to a trie node are stored contiguously next to each other. 
The start offset of each trie node cluster is maintained in a header section within the partition. 
Given this organization, the system loads a given data partition and  efficiently 
accesses the records belonging to a specific trie node. Each of these records (i.e., raw data series object) is compared against the query object using the ED distance (Def.~\ref{def:ed}) 
for accurate distance calculations. Finally, the scores are ranked for the final top K selection.

%===========================================================
%===========================================================

\section{Experimental Study}\label{sec:exper}

\subsection{Experimental Setup}
\label{sec:exp_setup}

\textbf{Cluster Configurations.} 
As a proof of concept, CLIMBER's prototype is realized on top of Apache Spark~\cite{zaharia2010spark}.
All introduced data structures and algorithms are built from scratch in Scala. 
All experiments are conducted on a cluster consisting of two nodes, each node has $56$ Intel@Xeon E5-2690 CPUs, $512GB$ RAM, and $8TB$ SATA hard drive. It runs Ubuntu 16.04.3 LTS with Apache Spark-$2.4.5$. The total memory available for user's data is around $850GB$
HDFS~\cite{shvachko2010hadoop}. 
The code for re-producibility is available at~\cite{climberCode}.

\textbf{Baseline Techniques.} Three distributed baseline techniques are included in the evaluation and compared 
against CLIMBER.  {\underline{\em Distributed Sequential Scan (Dss)}
 is the vanilla full scan solution that scans all data partitions in parallel to generate the exact answer 
 set (i.e., the ground truth) for the kNN queries. 
{\underline{\em TARDIS}}~\cite{zhang2019tardis} and {\underline{\em DPiSAX}}~\cite{yagoubi2017dpisax} are the state-of-the-art disk-based iSAX  
indexing techniques for distributed data series processing.  
They both create a global main-memory index structure and use it for 
re-partitioning the data and creating local indexes. They both 
support kNN approximate queries on big data series. 
Moreover, in Section~\ref{sec:exp_memory}, 
we extend our comparison to two recent distributed main-memory systems, 
namely Odyssey~\cite{Odyssey147779087} and 
ParlayANN-HNSW~\cite{1014562753538475}.

\textbf{CLIMBER Variations.}
Three variations of CLIMBER are evaluated.
As introduced in Section~\ref{sec:query}, 
{\em{\underline{CLIMBER-kNN}}} is the search algorithm presented 
in Algorithm~\ref{alg:kNN}. It targets only one trie node (and if needed it expands the search within the same partition). 
{\em{\underline{CLIMBER-kNN-Adaptive-2X}}} and 
{\em{\underline{CLIMBER-kNN-Adaptive-4X}}} are two variations of the 
adaptive algorithm in Section~\ref{sec:query}. These variations 
adaptively expand  the search to trie nodes 
in more than one partition  when  needed. 
The $2X$ and $4X$ suffixes indicate that these algorithms are 
capped to go after twice and four times the number of partitions compared to  CLIMBER-kNN, respectively.

\textbf{Real-World and Benchmark Datasets.} We use four datasets from different domains for the evaluation. The sizes of the datasets  range from 200 GB to 1.0 TB.
\textit{\underline{RandomWalk Benchmark}} is extensively used as benchmark for data series index in other projects\cite{camerra2010isax,ding2008querying,lin2007experiencing,shieh2008sax}. 
This dataset contains up to $1$ \textit{billion} data series, each having $256$ points.
\textit{\underline{Texmex Corpus}} \cite{jegou2011product}  is an image dataset which contains $1$ \textit{billion} SIFT feature vectors of $128$ points each. 
\textit{\underline{DNA}} \cite{uscs} dataset contains assembly of the human genome collected from $2000$ to $2013$. Each DNA string is divided into subsequences and then converted into data series as in \cite{camerra2010isax}. Each record consists of $192$ points.
\textit{\underline{Seizure EEG}} dataset contains records from dogs and humans with naturally occurring epilepsy. The data is sampled from 16 electrodes at 400 Hz, and recorded voltages were referenced to the group average. Each EEG record is split into $256$ points as a data series object.
For the query generation on each dataset, the query objects are randomly selected from the entire dataset. 
The query-related results are reported as the average over 50 queries.

\textbf{Default Parameters.}
Unless otherwise specified, parameters are set to default values.
These values are then  varied in specific experiments, as indicated. 
For example, the number of pivots in the system is set to 200, the pivot prefix length is set to 10, and the number of similar objects from a query (K) is set to 500. The default CLIMBER variation is set to {CLIMBER-kNN-Adaptive-4X}.

\subsection{CLIMBER Query and Index Evaluation.}
\label{sec:exp_CLIMBER}

In Figure~\ref{fig:Exp1}, we evaluate the query performance of CLIMBER against the three baseline techniques. 
For evaluation, we utilize two  metrics that are widely used in  other related works~\cite{park2015neighbor, nourachainlink, zhang2019tardis, yagoubi2017dpisax}. 
Namely, the query time is measured as the wall clock time from start to finish, 
which  incorporates as a dominant factor the number of partitions touched. 
%%%
While the query accuracy is measured using the standard recall metric as defined in Def.~\ref{def:knnapprox}.

In Figures~\ref{fig:Exp1}(a) and~(b), each dataset size is set to 200 GB and K is set to 500. 
Dss, which is the full-scan brute-force solution generates the exact answer set (i.e., recall = 1.0), 
however,  its execution time is prohibitively high and impractical to use. 
The other three algorithms are competitively very similar in their query execution time, while CLIMBER achieves between 25\% to 35\% higher accuracy. 
Reaching the 75\%-80\% range in recall is remarkable and unmatched in the existing literature given the large scale of the data. 

In Figures~\ref{fig:Exp1}(c) and~(d), we present the same two metrics while varying the dataset sizes from 200 GB to 1.0 TB using the RandomWalk benchmark. 
In some cases, CLIMBER encounters an additional query processing overhead between 5\% to 25\% compared to the baselines. 
This is due two reasons: (1)~CLIMBER needs to linearly search for the best matching groups in its index list, whereas both TARDIS and DPiSAX navigate a root-to-leaf path in an iSAX search tree, 
And (2)~For some queries, CLIMBER could {\em adaptively} process 
more than one data partition 
while the baseline techniques are restricted to a single partition.

Despite this extra overhead in query processing, which is potentially negligible to applications, CLIMBER shows consistent superiority in  
the recall metric. 
We contribute such improvement to two main reasons:
(1)~{\em effective multi-partition search:}
    CLIMBER's ranking of the trie nodes during the global index search enables it to intelligently expand the search across few partitions to acquire better candidates. In contrast, iSAX-based systems (TARDIS and DPiSAX) constraint their search to a single partition. 
And (2)~{\em more similarity-preserving  representation:}
    The iSAX representation is a kind of two-dimensional representation (x-axis represents the PAA segments and the y-axis represents the iSAX binary encoding) (Figure~\ref{fig:sax-isax-rep}). Both dimensions need to be kept short to preserve the iSAX tree scalability to big data (refer to Section~\ref{sec:art-limitations}). 
    In contrast, CLIMBER's P$^4$ dual signature operates directly on the PAA representation (and hence bypasses the information loss due to the iSAX encoding), and transfers this representation to a much higher space (i.e., the pivots space).
    Even after the reduction to the pivot-prefix space, the P$^4$ representation remains more similarity-preserving than iSAX.

In Figure~\ref{fig:Exp2}, we illustrate the index building overhead  for the different algorithms. Dss does not build an index, and thus is not included in   this set of experiments. 
The index construction time (Figures~\ref{fig:Exp2}(a) and (c)) includes all the steps  presented in Figure~\ref{fig:data-flow}, i.e., sampling, index skeleton construction, and data re-distribution. The other techniques TARDIS and DPiSAX also have similar steps. As can be observed in these two figures, DPiSAX takes the longest time to construct its index due to inefficient updates to its data structures while building the index.   CLIMBER involves a slightly higher construction time compared to TARDIS, because its pivot-based conversions and comparisons are more expensive compared to TARDIS's iSAX operations.
It is most important to note though that all three systems increase linearly as the dataset size increases as shown in 
Figure~\ref{fig:Exp2}(c).

The global index  (Figures~\ref{fig:Exp2}(b) and (d)) represents the index skeleton that is kept in the master node. 
As depicted in the figures, the global index size in the three systems is very tiny and fits easily in main memory. 
TARDIS has the largest size because its sigTree is a wide n-ary tree and its splitting algorithm creates 2 to 3 times more nodes compared to DPiSAX. 
CLIMBER, on the other hand, maintains two data structures, the groups list, and the forest of trie trees under these groups (Figure~\ref{fig:trie}).

\begin{figure}[t]
  \centering
  \includegraphics[width=1.0\columnwidth]{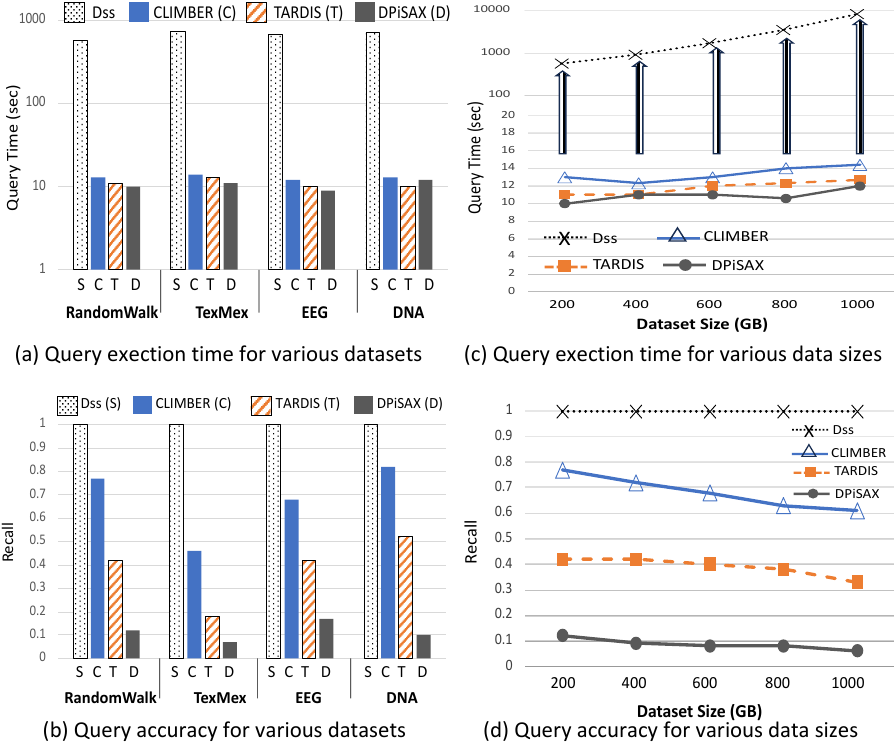}
  \caption{
  {\small{Evaluation of query execution. (a) and (b) illustrate the execution time and accuracy, respectively, for various datasets and algorithms (Dataset size = 200GB, K = 500). (c) and (d) illustrate the same for various data sizes and algorithms (Dataset = RandomWalk, K = 500). }}}
  \label{fig:Exp1}
\end{figure}

\begin{figure}[t]
  \centering
  \includegraphics[width=1.0\columnwidth]{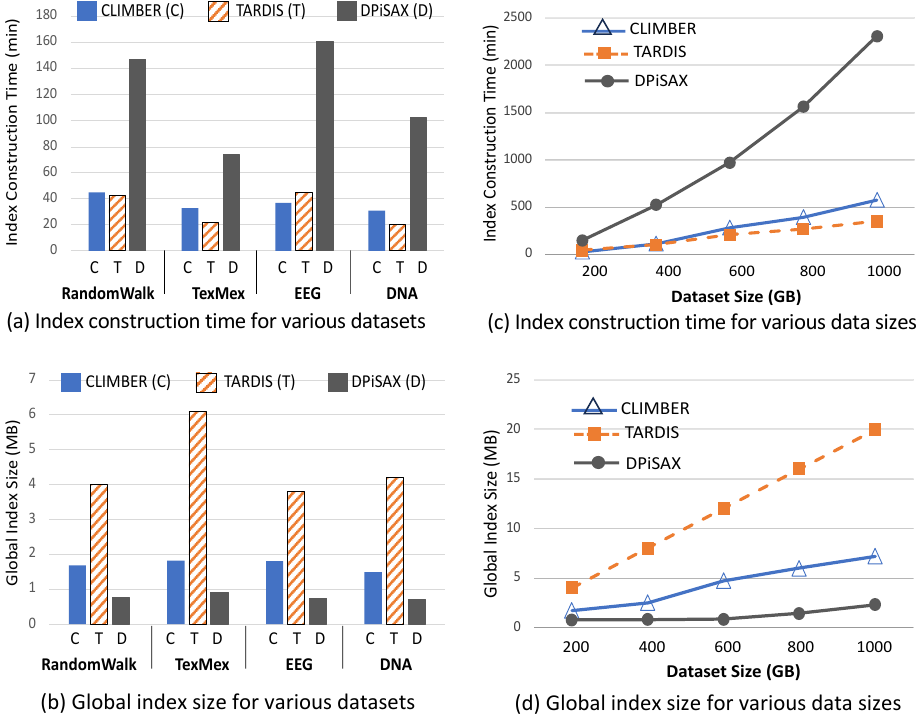}
  \caption{
  {\small{Evaluation of index building. (a) and (b) illustrate the construction time and index size, respectively, for various datasets and algorithms (Dataset size = 200GB). (c) and (d) illustrate the same for various data sizes and algorithms (Dataset = RandomWalk). }}}
  \label{fig:Exp2}
  \vspace{-3mm}
\end{figure}

\begin{figure}[t]
  \centering
  \includegraphics[width=1.0\columnwidth]{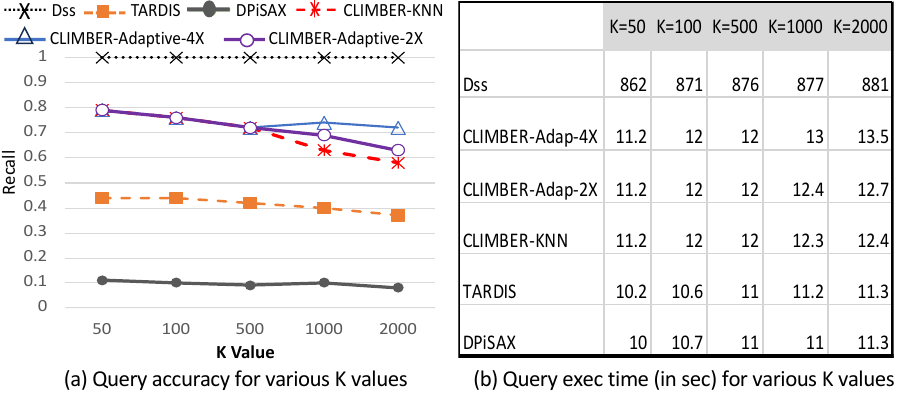}
  \caption{
  {\small{Evaluation of query execution under various K values. (a) illustrates the accuracy for various algorithms while (b) illustrates the execution time. (Dataset = RandomWalk, size = 400GB).}}}
  \label{fig:Exp3}
    \vspace{-2mm}
\end{figure}

In Figure~\ref{fig:Exp3}, we investigate the impact of varying the query answer size (K) on the performance of the different algorithms. Since this experiment stress tests larger K values, we included  the three variations of CLIMBER in this
evaluation. 
The key observations from Figure~\ref{fig:Exp3}(a) are that: 
(1)~Among the approximation algorithms, CLIMBER remains the superior one under all K values. 
(2)~Under small K values (50 to 500), the three CLIMBER variations exhibit the same performance because the adaptive algorithms behave in the same exact way as CLIMBER-kNN as long as the target trie node contains more than the requested K. 
(3)~Under larger K values, some queries reach trie nodes that contain less than the requested K. 
In this case, we observe that the adaptive variations of CLIMBER become more robust as
they can identify more than one partition to search in.
More results on this phenomenon are discussed in Section~\ref{sec:exp_variations}.

The table in Figure~\ref{fig:Exp3}(b) illustrates the query execution time for the various algorithms under different K values. All approximation algorithms are in the same ballpark. 
The CLIMBER adaptive variations become slightly higher than CLIMBER-kNN because they tend to access few  extra partitions to seek higher quality results.

\subsection{Ablation Study of CLIMBER Variations.}
\label{sec:exp_variations}

In the following set of experiments, we investigate different design parameters within CLIMBER and study further the three variations of our system. 

In Figure~\ref{fig:Exp4}, we vary the number of pivots in the system (the default value is 200). This parameter impacts all aspects of the system at different degrees. For example, the impact on the index construction time 
is illustrated in Figure~\ref{fig:Exp4}(a). In this experiment, we zoom  into the three construction phases highlighted in the figure. The impact on the skeleton building phase 
is very minimal, mainly because this phase operates on a small sample, and also the truncation of signatures to the fixed prefix length (the default is 10) masks the increase in the number of pivots. However, the impact is more noticeable when operating on the entire datasets - in particular,
for data conversion and re-distribution both  part of Step 4 in Figure~\ref{fig:data-flow}).  
As expected, the comparison operations for measuring distances takes longer as the number of pivots increases. 

In Figure~\ref{fig:Exp4}(b), we illustrate the impact of the number of pivots on the query accuracy. We studied the impact on the four datasets.  They all exhibit similar trend. Basically, too few pivots results in data groups having too coarse granularities with less similarity preservation. In contrast,  too many pivots causes the curse of dimensionality problem where distances lose their true meaning. 
Empirically, we found that the range from 150 to 250 pivots yields the best results. 

We also studied the impact of the number of pivots on
~the query execution time. We found the impact negligible since it is only one object (the query object) to compare against the {\em n} pivots to compute the full-length pivot representation. Once truncated based on the prefix length, all further query processing steps remain the same. 
We also studied their impact on 
the global index size. Given  this size is tiny as reported in Figure~\ref{fig:Exp2},  we observed around 10\% to 20\% increase in 
this index size in part because the number of distinct pivot prefixes increases 
as the pool of total pivots increases.

In Figure~\ref{fig:Exp5}, we dig deeper into the differences among the three CLIMBER variations. The CLIMBER's adaptive variations start to outperform CLIMBER-kNN when the requested query answer size (K) becomes larger than the capacity of the trie node found by the search algorithm (refer to Algorithm~\ref{alg:kNN}).   
To stress test this case, given a query object, we first let the search algorithm find this target trie node.
We refer to its capacity as {\em m}. We then vary K to be range from m to 10m as illustrated in Figure~\ref{fig:Exp5}(a) (x axis). On the y-axis, we show the 
percentage of recall improvement that can be achieved from leveraging the adaptive algorithms. The numbers inside the bubbles represent the absolute recall achieved from CLIMBER-kNN. As can be observed, even when K = 2m, around 5\% improvement is measured. Yet, in the extreme cases, more than 40\% can be achieved. 

In the experiment conducted in Figure~\ref{fig:Exp5}(b), we wanted to understand the trade-offs between narrowing the query search to a specific trie node(s) and few partitions (i.e., limiting the data examined to answer a query) and scanning all data groups whose OD distance to the query object is the smallest (i.e., stopping at Line 6 in Algorithm~\ref{alg:kNN}). We refer to the latter as the {\em OD-Smallest} search algorithm. Figure~\ref{fig:Exp5}(b) illustrates the relative score of the {\em OD-Smallest} algorithm 
divided by the corresponding score from each of CLIMBER's three variations. We focused on two measures; the amount of accessed data to answer the query, and the recall. 

In summary, since the {\em OD-Smallest} algorithm scans more data, it  achieves a recall improvement. However, even under scanning 6x to 7x more data partitions, the 
improvement in recall is less than 10\% when compared to CLIMBER's default variation, namely, CLIMBER-kNN-Adaptive-4X. This  indicates that the proposed trie-based partitioning is effective in clustering high-quality data such that the search 
algorithms can process significantly less amount of data at query time without sacrificing too much of the query accuracy.

\begin{figure}[t]
  \centering
  \includegraphics[width=1.0\columnwidth]{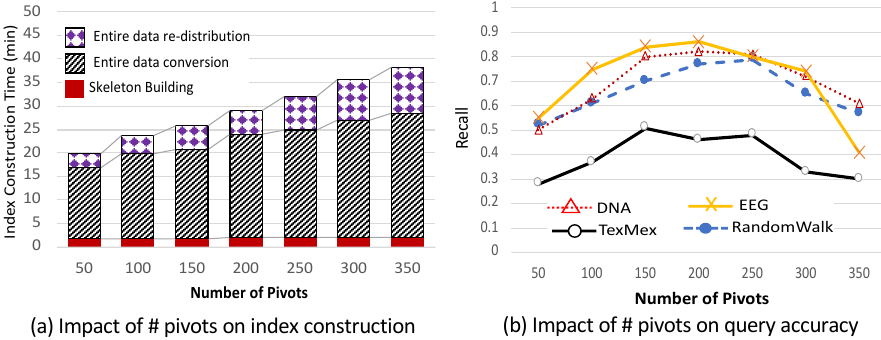}
  \caption{
  {\small{Evaluation of the number of pivots. (a) illustrates the impact on index construction time divided into its main phases (Dataset = RandomWalk, size = 200GB). (b) illustrates the impact on the query accuracy for the various datasets (Size = 200GB, K = 500).}}
  }
  \label{fig:Exp4}
\end{figure}

\begin{figure}[t]
  \centering
  \includegraphics[width=1.0\columnwidth]{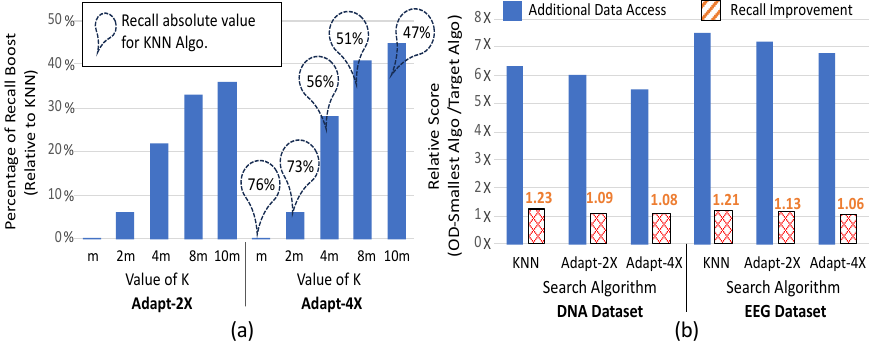}
  \caption{
  {\small{
  Evaluation of CLIMBER's query processing variations. 
  (a) compares the adaptive variations to the non-adaptive kNN variation.
  (b) compares the three variations to the OD-Smallest Algorithm. }}}
  \label{fig:Exp5}
    \vspace{-2mm}
\end{figure}

%================================

\begin{figure}[t]
  \centering
  \includegraphics[width=0.8\columnwidth]{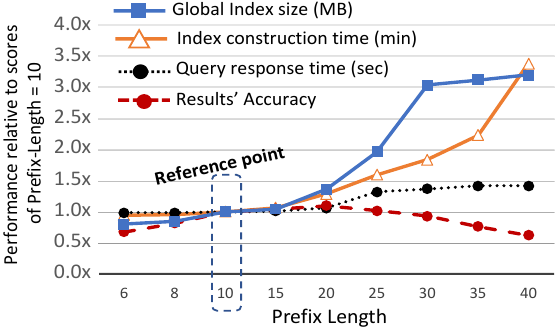}
  \caption{
  {\small{
  Evaluation of the prefix length (Dataset = RandomWalk, Size = 400GB, K = 500). The performance of the different metrics is relative to the scores of the default prefix-length = 10 (the absolute scores are: global index size = 2.5MB, index construction time = 91 min, query response time = 12.3 sec, and query accuracy = 0.71).
  }}
  }
  \label{fig:Exp5}
\end{figure}

In Figure~\ref{fig:Exp5}, we study the impact of the prefix length parameter on the system's performance. The default setting for this parameter in all other experiments is 10. In this experiment, we vary the length from 6 to 40, and present on the y-axis the  performance of the four highlighted metrics relative to that of prefix-length = 10. When the length is too short (e.g., 6 or 8), the results' accuracy decreases rapidly. This is because the objects' representations become too coarse, and the group/partition on which the search algorithm lands$-$assuming it contains the best matches$-$may contain low-quality matches to 
the query object. 
On the other hand, increasing the prefix length clearly results in increasing several metrics. The global index size increases because the number of distinct permutations of the prefixes increase, which increases the number of groups and trie nodes in the global index (refer to Fig.~\ref{fig:trie}). However, it stabilizes at a certain point due to the safe guards in Algorithm~\ref{alg:groupAssign} that prevents creating too nearby or tiny groups.

Another metric that has clearly increased as the prefix length increases is the index construction time. This is primarily due to Step 4 in Figure~\ref{fig:data-flow} in which the entire dataset is re-distributed based on the objects' prefix representations. The query response time is relatively stable under prefix length changes although it showed increase starting from length = 25. We attribute this to the fact that when the length gets too long (25 and above) the space becomes highly fragmented and the trie-node on which the search algorithm lands may not contain enough objects that covers K, and hence the algorithm is forced to search for and load more partitions. 
This observation also explains the decrease in the results' accuracy at length 25 and above--despite an increase to 73\% and 74\% for lengths 15 and 20, respectively.  
Basically, CLIMBER-kNN-Adaptive-4X caps the number of loaded partitions to 4 while in a highly fragmented space, tiny trie-nodes containing high-quality matches  could get scattered over many partitions.
Overall, the results indicate that between 10 and 20 is an ideal choice for this parameter.

\subsection{{CLIMBER vs. Memory-Based Systems.}}
\label{sec:exp_memory}

In this section, we present a comparison among three 
distinct approaches for answering kNN similarity search queries with 
the objective of highlighting their major pros and cons. 
The three systems under comparison are: 
CLIMBER, Odyssey~\cite{Odyssey147779087}, 
and ParlayANN-HNSW (ParlayANN for short)~\cite{1014562753538475}.  
%CLIMBER is our distributed disk-based system for approximate answering of ad-hoc kNN queries. 
Odyssey is a distributed memory-based system for exact answering of batch kNN queries. 
It relies on iSAX trees~\cite{camerra2010isax} for constructing its index. 
Odyssey's main strength lies in the scheduling and load balancing mechanisms it offers for 
efficiently executing 100s of queries concurrently. 
ParlayANN is a single-node multi-core memory-based system for approximate answering of ad-hoc kNN queries.
ParlayANN is a based on graph similarity search.

In this experiment, we use RandomWalk dataset of different sizes. All default out-of-the-box settings of Odyssey and ParlayANN remain intact with the exception of setting K = 500.  
Odyssey is configured with Slurm framework to run on the cluster's two nodes, while 
ParlayANN runs on only a single node. 
We measure the three performance metrics highlighted in Table~\ref{tab:v2}, namely 
{\em Index Construction Time (I.C.T), Query Response Time (Q.R.T), and Result Recall (R.R)}.

The key insights from the results in Table~\ref{tab:v2} are:
(1)~CLIMBER achieves its objective of scaling 
    to large datasets, which other main-memory systems fail to support, while retaining a consistent and acceptable query response time (below 20 sec.) and 
    relatively good result's accuracy (recall is mostly 60\% and above). 
(2)~Certainly, CLIMBER is slower than Odyssey as long as the data fits in memory. 
    For example, data re-distribution and writing to the distributed file system (along with all involved replication) results in 2x to 3x higher I.C.T metric for CLIMBER. 
    Similarly, Q.R.T is slower in CLIMBER since it loads few HDFS partitions for answering the query while Odyssey performs main-memory operations. 
And (3)~Although, the query response time (Q.R.T) for the main-memory systems (Odyssey \& ParlayANN) is better than CLIMBER, in these systems, if the data is not readily available in memory, then Q.R.T will radically jump  to several minutes at query time for re-loading the data and indexes into memory. 
In contrast,  CLIMBER's  Q.R.T is consistently below 20 sec.

The results also show that graph-based methods are very promising with respect to the 
query throughput and results' accuracy metrics. 
Nevertheless, scaling these techniques to operate on big data in highly distributed settings remains 
a big challenge especially during the graph construction phase which took more than 10 hours for a relatively mid-size dataset.

\begin{table}[t]
    {\small{
    \centering
    \begin{tabular*}{0.95\columnwidth}[b]{l|l|l|l|l}
        \hline
         Size(GB) & Metric & CLIMBER & Odyssey & ParlayANN$^*$   \\
          \hline
         \multirow{3}{*}{200} & I.C.T & 27 & 14 & 218     \\
                              & Q.R.T & 13 & 0.7 &  0.14  \\    %%0.34~\|~
                              & R.R & 0.77 & 1.0 & 0.92    \\
         \hline
         \multirow{3}{*}{400} & I.C.T & 91 & 48.3 & 776 \\
                              & Q.R.T & 12.3 & 1.4 & 0.21  \\   %%0.46~\|~
                              & R.R & 0.71 & 1.0 & 0.92  \\
         \hline
         \multirow{3}{*}{600} & I.C.T & 280 & 67.3 & X \\
                              & Q.R.T & 13.1 & 1.6 & X  \\
                              & R.R & 0.68 & 1.0 & X  \\
         \hline
         \multirow{3}{*}{800} & I.C.T & 390 & 112.8 & X \\
                              & Q.R.T & 14 & 2.0 & X  \\
                              & R.R & 0.63 & 1.0 & X  \\
         \hline
         \multirow{3}{*}{1000} & I.C.T & 576 & X & X \\
                              & Q.R.T & 14.4 & X & X  \\
                              & R.R & 0.62 & X & X  \\
         \hline
         \multirow{3}{*}{1500} & I.C.T & 875 & X & X \\
                              & Q.R.T & 17.2 & X & X  \\
                              & R.R & 0.56 & X & X  \\
         \hline
         \multicolumn{5}{l}{I.C.T: Index Construction Time (mins).}  \\
         \multicolumn{5}{l}{Q.R.T: Query Response Time (secs).}  \\
         \multicolumn{5}{l}{R.R: Results' Recall (lowest = 0.0, and highest (best) = 1.0).}  \\
         \multicolumn{5}{l}{$^*$For ParlayANN, the entire dataset in contained in one node.}  \\

        \end{tabular*}
    }}
      \caption{\small{Comparison with In-Memory Systems.}}
  \label{tab:v2}
\end{table}

\section{Conclusion}
\label{sec:conclusion}
In conclusion, in this work, we have proposed  a novel distributed framework for feature extraction, indexing, and approximate kNN query processing over big data series called CLIMBER.  Our CLIMBER solution is distinct from  state-of-the-art distributed techniques in its ability to achieve both  scalability and high query accuracy within the same system. 
For CLIMBER, we design a novel pivot-based dual representation for data series objects, new distance metrics that suit this unique dual representation, 
innovative data partitioning and algorithms 
for our proposed index structures, and efficient yet effective  kNN query processing algorithms.  
Our comprehensive experimental study over numerous
real-world and benchmark data sts 
 demonstrate the superiority of CLIMBER over state-of-the-art solutions.

\small
\bibliographystyle{abbrv}
\bibliography{paper}

\begin{thebibliography}{10}

\bibitem{climberCode}
{CLIMBER Source Code}.
\newblock \url{https://github.com/lzhang6/climber.git }.

\bibitem{abanda2019review}
A.~Abanda, U.~Mori, and J.~A. Lozano.
\newblock A review on distance based time series classification.
\newblock {\em Data Mining and Knowledge Discovery}, 33(2):378--412, 2019.

\bibitem{abraham1989outlier}
B.~Abraham and A.~Chuang.
\newblock Outlier detection and time series modeling.
\newblock {\em Technometrics}, 31(2):241--248, 1989.

\bibitem{aghabozorgi2015time}
S.~Aghabozorgi, A.~S. Shirkhorshidi, and T.~Y. Wah.
\newblock Time-series clustering--a decade review.
\newblock {\em Information Systems}, 53:16--38, 2015.

\bibitem{nourachainlink}
N.~Alghamdi, L.~Zhang, M.~Y. Eltabakh, and E.~A. Rundensteiner.
\newblock Chainlink: Indexing big time series data for long subsequence matching.
\newblock In {\em ICDE}. IEEE, 2020.

\bibitem{aljawarneh2016similarity}
S.~Aljawarneh, V.~Radhakrishna, P.~V. Kumar, and V.~Janaki.
\newblock A similarity measure for temporal pattern discovery in time series data generated by iot.
\newblock In {\em 2016 International conference on engineering \& MIS (ICEMIS)}, pages 1--4. IEEE, 2016.

\bibitem{amato2008approximate}
G.~Amato and P.~Savino.
\newblock Approximate similarity search in metric spaces using inverted files.
\newblock In {\em Proceedings of the 3rd international conference on Scalable information systems}, pages 1--10, 2008.

\bibitem{arge2008priority}
L.~Arge, M.~D. Berg, H.~Haverkort, and K.~Yi.
\newblock The priority r-tree: A practically efficient and worst-case optimal r-tree.
\newblock {\em TALG}, 4(1):1--30, 2008.

\bibitem{arora2018hd}
A.~Arora, S.~Sinha, P.~Kumar, and A.~Bhattacharya.
\newblock Hd-index: Pushing the scalability-accuracy boundary for approximate knn search in high-dimensional spaces.
\newblock {\em Proc. VLDB Endow.}, 11(8), 2018.

\bibitem{beckmann1990r}
N.~Beckmann, H.-P. Kriegel, R.~Schneider, and B.~Seeger.
\newblock The r*-tree: an efficient and robust access method for points and rectangles.
\newblock In {\em SIGMOD}, volume~19, pages 322--331. ACM, 1990.

\bibitem{bellman1966dynamic}
R.~Bellman.
\newblock Dynamic programming.
\newblock {\em Science}, 153(3731):34--37, 1966.

\bibitem{camerra2010isax}
A.~Camerra, T.~Palpanas, J.~Shieh, and E.~Keogh.
\newblock isax 2.0: Indexing and mining one billion time series.
\newblock In {\em ICDM}, pages 58--67. IEEE, 2010.

\bibitem{carrington2005models}
P.~J. Carrington, J.~Scott, and S.~Wasserman.
\newblock {\em Models and methods in social network analysis}, volume~28.
\newblock Cambridge university press, 2005.

\bibitem{chan1999efficient}
K.-P. Chan and A.~W.-C. Fu.
\newblock Efficient time series matching by wavelets.
\newblock In {\em ICDE}, pages 126--133. IEEE, 1999.

\bibitem{chan2004time}
N.~H. Chan.
\newblock {\em Time series: applications to finance}, volume 487.
\newblock John Wiley \& Sons, 2004.

\bibitem{Odyssey147779087}
M.~Chatzakis, P.~Fatourou, E.~Kosmas, T.~Palpanas, and B.~Peng.
\newblock Odyssey: A journey in the land of distributed data series similarity search.
\newblock {\em Proc. VLDB Endow.}, 16(5):1140–1153, jan 2023.

\bibitem{chavez2005proximity}
E.~Ch{\'a}vez, K.~Figueroa, and G.~Navarro.
\newblock Proximity searching in high dimensional spaces with a proximity preserving order.
\newblock In {\em Mexican International Conference on Artificial Intelligence}, pages 405--414. Springer, 2005.

\bibitem{cook2019anomaly}
A.~A. Cook, G.~M{\i}s{\i}rl{\i}, and Z.~Fan.
\newblock Anomaly detection for iot time-series data: A survey.
\newblock {\em IEEE Internet of Things Journal}, 7(7):6481--6494, 2019.

\bibitem{PostgreSQL}
{Database}.
\newblock Postgresql: The world's most advanced open source relational database.
\newblock \url{https://www.postgresql.org}.

\bibitem{delorme2016bin}
M.~Delorme, M.~Iori, and S.~Martello.
\newblock Bin packing and cutting stock problems: Mathematical models and exact algorithms.
\newblock {\em EJOR}, 2016.

\bibitem{ding2008querying}
H.~Ding, G.~Trajcevski, P.~Scheuermann, X.~Wang, and E.~Keogh.
\newblock Querying and mining of time series data: experimental comparison of representations and distance measures.
\newblock {\em PVLDB}, 1, 2008.

\bibitem{echihabi2019return}
K.~Echihabi, K.~Zoumpatianos, T.~Palpanas, and H.~Benbrahim.
\newblock Return of the lernaean hydra: experimental evaluation of data series approximate similarity search.
\newblock {\em PVLDB}, 13(3):403--420, 2019.

\bibitem{esling2012time}
P.~Esling and C.~Agon.
\newblock Time-series data mining.
\newblock {\em ACM Computing Surveys (CSUR)}, 45(1):12, 2012.

\bibitem{esuli2012use}
A.~Esuli.
\newblock Use of permutation prefixes for efficient and scalable approximate similarity search.
\newblock {\em Information Processing \& Management}, 48(5):889--902, 2012.

\bibitem{faloutsos1994fast}
C.~Faloutsos, M.~Ranganathan, and Y.~Manolopoulos.
\newblock {\em Fast subsequence match in time-series databases}.
\newblock ACM, 1994.

\bibitem{Dolam23}
{Github}.
\newblock Ai2 dolma: 3 trillion token open corpus for language model pretraining.
\newblock \url{https://github.com/allenai/dolma}.

\bibitem{PyNNDescent}
{Github}.
\newblock Pynndescent for fast approximate nearest neighbors.
\newblock \url{https://github.com/lmcinnes/pynndescent}.

\bibitem{RedPajama}
{Github}.
\newblock Redpajama-data-v2: an open dataset with 30 trillion tokens for training large language models.
\newblock \url{https://github.com/togethercomputer/RedPajama-Data}.

\bibitem{gonzalez2008effective}
E.~C. Gonzalez, K.~Figueroa, and G.~Navarro.
\newblock Effective proximity retrieval by ordering permutations.
\newblock {\em IEEE Transactions on Pattern Analysis and Machine Intelligence}, 30(9):1647--1658, 2008.

\bibitem{guttman1984r}
A.~Guttman.
\newblock R-trees: A dynamic index structure for spatial searching.
\newblock In {\em Proceedings of the 1984 ACM SIGMOD international conference on Management of data}, pages 47--57, 1984.

\bibitem{jegou2011product}
H.~Jegou, M.~Douze, and C.~Schmid.
\newblock Product quantization for nearest neighbor search.
\newblock {\em TPAMI}, pages 117--128, 2011.

\bibitem{jensen2017time}
S.~K. Jensen, T.~B. Pedersen, and C.~Thomsen.
\newblock Time series management systems: A survey.
\newblock {\em TKDE}, 29(11):2581--2600, 2017.

\bibitem{DBLP:journals/corr/JohnsonDJ17}
J.~Johnson, M.~Douze, and H.~J{\'{e}}gou.
\newblock Billion-scale similarity search with gpus.
\newblock {\em CoRR}, abs/1702.08734, 2017.

\bibitem{kamel1993hilbert}
I.~Kamel and C.~Faloutsos.
\newblock Hilbert r-tree: An improved r-tree using fractals.
\newblock Technical report, 1993.

\bibitem{keogh2001dimensionality}
E.~Keogh, K.~Chakrabarti, M.~Pazzani, and S.~Mehrotra.
\newblock Dimensionality reduction for fast similarity search in large time series databases.
\newblock {\em KAIS}, 3:263--286, 2001.

\bibitem{kondylakis2019coconut}
H.~Kondylakis, N.~Dayan, K.~Zoumpatianos, and T.~Palpanas.
\newblock Coconut palm: Static and streaming data series exploration now in your palm.
\newblock In {\em ACM SIGMOD}, 2019.

\bibitem{10.172749}
R.~Kumar and S.~Vassilvitskii.
\newblock Generalized distances between rankings.
\newblock In {\em Proceedings of the 19th International Conference on World Wide Web}, page 571–580, 2010.

\bibitem{gh87287}
J.~Leskovec, A.~Rajaraman, and J.~D. Ullman.
\newblock Mining of massive datasets 3rd edition.
\newblock In {\em Cambridge university press}, 2020.

\bibitem{lin2007experiencing}
J.~Lin, E.~Keogh, L.~Wei, and S.~Lonardi.
\newblock Experiencing sax: a novel symbolic representation of time series.
\newblock {\em DMKD}, 2007.

\bibitem{DBLP:conf/nips/LuoS16}
C.~Luo and A.~Shrivastava.
\newblock {SSH} (sketch, shingle, {\&} hash) for indexing massive-scale time series.
\newblock In {\em Proceedings of the {NIPS} 2016 Time Series Workshop}, volume~55 of {\em {JMLR} Workshop and Conference Proceedings}, pages 38--58, 2016.

\bibitem{10.110ww89473}
Y.~A. Malkov and D.~A. Yashunin.
\newblock Efficient and robust approximate nearest neighbor search using hierarchical navigable small world graphs.
\newblock {\em IEEE Trans. Pattern Anal. Mach. Intell.}, 42(4):824–836, 2020.

\bibitem{1014562753538475}
M.~D. Manohar, Z.~Shen, G.~Blelloch, L.~Dhulipala, Y.~Gu, H.~V. Simhadri, and Y.~Sun.
\newblock Parlayann: Scalable and deterministic parallel graph-based approximate nearest neighbor search algorithms.
\newblock In {\em Proceedings of the 29th ACM SIGPLAN Annual Symposium on Principles and Practice of Parallel Programming}, page 270–285, 2024.

\bibitem{mottin2019exploring}
D.~Mottin, M.~Lissandrini, Y.~Velegrakis, and T.~Palpanas.
\newblock Exploring the data wilderness through examples.
\newblock In {\em SIGMOD}. ACM, 2019.

\bibitem{novak2011metric}
D.~Novak, M.~Batko, and P.~Zezula.
\newblock Metric index: An efficient and scalable solution for precise and approximate similarity search.
\newblock {\em Information Systems}, 36(4):721--733, 2011.

\bibitem{novak2016ppp}
D.~Novak and P.~Zezula.
\newblock Ppp-codes for large-scale similarity searching.
\newblock In {\em Transactions on Large-Scale Data-and Knowledge-Centered Systems XXIV}, pages 61--87. Springer, 2016.

\bibitem{palpanas2016big}
T.~Palpanas.
\newblock Big sequence management: A glimpse of the past, the present, and the future.
\newblock In {\em SOFSEM}, pages 63--80. Springer, 2016.

\bibitem{palpanas2017parallel}
T.~Palpanas.
\newblock The parallel and distributed future of data series mining.
\newblock In {\em HPCS}, pages 916--920. IEEE, 2017.

\bibitem{isaxfamily}
T.~Palpanas.
\newblock Evolution of a data series index.
\newblock In {\em ISIP}. Springer, 2019.

\bibitem{park2015neighbor}
Y.~Park, M.~Cafarella, and B.~Mozafari.
\newblock Neighbor-sensitive hashing.
\newblock {\em Proceedings of the VLDB Endowment}, 9(3):144--155, 2015.

\bibitem{9101877Messi}
B.~Peng, P.~Fatourou, and T.~Palpanas.
\newblock Messi: In-memory data series indexing.
\newblock In {\em 2020 IEEE 36th International Conference on Data Engineering (ICDE)}, pages 337--348, 2020.

\bibitem{10.1Messi00677-2}
B.~Peng, P.~Fatourou, and T.~Palpanas.
\newblock Fast data series indexing for in-memory data.
\newblock {\em The VLDB Journal}, 30(6):1041–1067, jun 2021.

\bibitem{samet2006foundations}
H.~Samet.
\newblock {\em Foundations of multidimensional and metric data structures}.
\newblock Morgan Kaufmann, 2006.

\bibitem{sclove1983time}
S.~L. Sclove.
\newblock Time-series segmentation: A model and a method.
\newblock {\em Information Sciences}, 29(1):7--25, 1983.

\bibitem{shieh2008sax}
J.~Shieh and E.~Keogh.
\newblock isax: indexing and mining terabyte sized time series.
\newblock In {\em SIGKDD}, pages 623--631. ACM, 2008.

\bibitem{shvachko2010hadoop}
K.~Shvachko, H.~Kuang, S.~Radia, R.~Chansler, et~al.
\newblock The hadoop distributed file system.
\newblock In {\em MSST}, volume~10, pages 1--10, 2010.

\bibitem{skala2009counting}
M.~Skala.
\newblock Counting distance permutations.
\newblock {\em Journal of Discrete Algorithms}, 7(1):49--61, 2009.

\bibitem{105555645803669368}
Z.~R. Struzik and A.~Siebes.
\newblock The haar wavelet transform in the time series similarity paradigm.
\newblock In {\em Proceedings of the Third European Conference on Principles of Data Mining and Knowledge Discovery}, PKDD '99, page 12–22, Berlin, Heidelberg, 1999. Springer-Verlag.

\bibitem{10.555555520}
S.~J. Subramanya, Devvrit, R.~Kadekodi, R.~Krishaswamy, and H.~V. Simhadri.
\newblock {\em DiskANN: fast accurate billion-point nearest neighbor search on a single node}.
\newblock 2019.

\bibitem{tellez2011succinct}
E.~S. Tellez, E.~Ch{\'a}vez, and G.~Navarro.
\newblock Succinct nearest neighbor search.
\newblock In {\em Proceedings of the Fourth International Conference on SImilarity Search and APplications}, pages 33--40, 2011.

\bibitem{torkamani2017survey}
S.~Torkamani and V.~Lohweg.
\newblock Survey on time series motif discovery.
\newblock {\em Wiley Interdisciplinary Reviews: Data Mining and Knowledge Discovery}, 7(2):e1199, 2017.

\bibitem{uscs}
UCSC.
\newblock https://genome.ucsc.edu/.

\bibitem{Chroma333}
{Vector DBs}.
\newblock Chroma: the ai-native open-source embedding database.
\newblock \url{https://www.trychroma.com}.

\bibitem{wei2006time}
W.~W. Wei.
\newblock Time series analysis.
\newblock In {\em The Oxford Handbook of Quantitative Methods in Psychology: Vol. 2}. 2006.

\bibitem{wu2019kv}
J.~Wu, P.~Wang, N.~Pan, C.~Wang, W.~Wang, and J.~Wang.
\newblock Kv-match: A subsequence matching approach supporting normalization and time warping.
\newblock In {\em ICDE}. IEEE, 2019.

\bibitem{yagoubi2017dpisax}
D.-E. Yagoubi, R.~Akbarinia, F.~Masseglia, and T.~Palpanas.
\newblock {DPiSAX: Massively Distributed Partitioned iSAX}.
\newblock In {\em ICDM}, pages 1135--1140. IEEE, 2017.

\bibitem{zaharia2010spark}
M.~Zaharia, M.~Chowdhury, M.~J. Franklin, S.~Shenker, and I.~Stoica.
\newblock Spark: Cluster computing with working sets.
\newblock {\em HotCloud}, 2010.

\bibitem{zhang2019tardis}
L.~Zhang, N.~Alghamdi, M.~Y. Eltabakh, and E.~A. Rundensteiner.
\newblock Tardis: Distributed indexing framework for big time series data.
\newblock In {\em ICDE}, pages 1202--1213. IEEE, 2019.

\bibitem{zhang2020big}
L.~Zhang, N.~Alghamdi, M.~Y. Eltabakh, and E.~A. Rundensteiner.
\newblock Big data series analytics using tardis and its exploitation in geospatial applications.
\newblock In {\em SIGMOD}, pages 2785--2788. ACM, 2020.

\bibitem{10.147777374}
B.~Zheng, X.~Zhao, L.~Weng, N.~Q.~V. Hung, H.~Liu, and C.~S. Jensen.
\newblock Pm-lsh: A fast and accurate lsh framework for high-dimensional approximate nn search.
\newblock {\em Proc. VLDB Endow.}, 13(5):643–655, jan 2020.

\bibitem{zoumpatianos2016ads}
K.~Zoumpatianos, S.~Idreos, and T.~Palpanas.
\newblock Ads: the adaptive data series index.
\newblock {\em VLDB}, 2016.

\end{thebibliography}

\end{document}